\documentclass[12pt]{article}
\usepackage[utf8]{inputenc}
\usepackage[a4paper, top=25mm, bottom=25mm, left=25mm, right=20mm]{geometry}
\usepackage{caption}
\usepackage{subcaption}
\usepackage[authoryear]{natbib}
\usepackage{subfiles}
\usepackage{booktabs}
\usepackage{blindtext}
\usepackage{tikz}
\usetikzlibrary{positioning}
\usetikzlibrary{calc}
\usepackage{float}
\usepackage{mathtools}
\usepackage{comment}
\usepackage{verbatim}
\usepackage{mdframed}
\usepackage{graphicx}
\usepackage{appendix}
\usepackage{chngcntr}
\usepackage{etoolbox}
\usepackage[official]{eurosym}
\usepackage{multirow}
\usepackage{amsthm}
\usepackage{color} 
\usepackage{xcolor}
\usepackage{empheq}
\usepackage{quoting}
\quotingsetup{font=small}
\usepackage{amssymb}
\usepackage{graphicx}
\usepackage{float}
\usepackage{amssymb}
\usepackage{amsmath}
\usepackage{systeme}
\usepackage[labelfont=bf]{caption}
\usepackage{fancyhdr}
\usepackage{siunitx}  
\usepackage{framed}
\usepackage{multirow}

\title{\huge \textbf{Bridging Distant Ideas: the Impact of AI on R\&D and Recombinant Innovation} \medskip
    \author{Emanuele Bazzichi$^{1,*}$, Massimo Riccaboni$^{1,2}$ and Fulvio Castellacci$^{3}$}
    \date{This version: 31st October 2026}\\}

\begin{document}

\maketitle


\begin{abstract}
\noindent We study how artificial intelligence (AI) affects firms' incentives to pursue incremental \textit{versus} radical knowledge recombinations. We develop a model of recombinant innovation embedded in a Schumpeterian quality-ladder framework, in which innovation arises from recombining ideas across varying distances in a knowledge space. R\&D consists of multiple tasks, a fraction of which can be performed by AI. AI facilitates access to distant knowledge domains, but at the same time it also increases the aggregate rate of creative destruction, shortening the monopoly duration that rewards radical innovations. We obtain three main results. First, the effect of increasing the share of AI-automated R\&D tasks is non-monotonic under specific conditions: firms initially target more radical innovations, but beyond a threshold they shift the focus toward incremental innovations. Second, with falling optimal distance, we may have increasing expected log-growth of knowledge due to higher innovation frequency or a hump-shaped growth. Finally, we identify a sufficient condition under which the target distance chosen in the decentralized equilibrium is sub-optimal from a social welfare point of view. \\ 
\end{abstract}

\noindent\rule{\textwidth}{0.6pt} 
$^{1}$ IMT School for Advanced Studies Lucca, Lucca, Italy.\\
$^2$ IUSS, University School for Advanced Studies of Pavia, Pavia, Italy.\\
$^3$ TIK Centre, University of Oslo, Oslo, Norway.\\
$^*$ Corresponding author, emanuele.bazzichi@imtlucca.it
\thispagestyle{empty}

\clearpage

\section{Introduction}
The central objective of scientific research is the production of new knowledge that improves our understanding of natural and social phenomena. Historically, advances in science have led to the discovery of new sources and forms of energy, new technologies and production methods, as well as a range of new products which increased life expectancy and welfare. By fueling industrialization, scientific progress has played a central role in raising living standards and promoting unprecedented economic growth.

\textit{But how is new knowledge created?} \textit{New ideas are mainly produced through the recombination of already existing knowledge in a cumulative and combinatorial process} \citep{weitzman1998}. One can see existing ideas as seeds that can be cross-pollinated to create new varieties of knowledge, in the same way one creates new plants transferring the pollen from the flower of one plant to another. This recombinant view of innovation has a long tradition in economics. \citet{schumpeter1934} identified innovation with ``new combinations'' of means of production carried out by entrepreneurs. Similarly, \citet{marx1894kapital} wrote of ``[...] new combinations, i.e., to increase the productivity of a given quantity of labour [...]''.

For example, Edison described how the idea of a domestic electric lighting network came to his mind by combining his own invention, the ``electric candle'' -- an early name for the light bulb -- with the already established idea of a gas distribution system \citep{weitzman1998}. A different example is provided by the Hubble Space Telescope. As its first images arrived on Earth blurred due to a problem of spherical aberration in the primary mirror, scientists applied existing deconvolution algorithms to denoise them. The operation was a success and these algorithms were subsequently adopted for early screening in medicine, saving many lives \citep{bonaventura2022}. These anecdotes highlight that new knowledge and innovations (mainly, but not uniquely) result from the combination of already established ideas, sometimes coming from quite different fields.

Fueled by the abundance of data and computing power, and advances in algorithms, Artificial Intelligence (AI) -- in particular, machine learning and Large Language Models (LLMs) -- has rapidly progressed since the 2010s and has begun to transform how scientists work and how discoveries are made. AI is being applied to analyze complex datasets, to formulate new theories and to solve problems that have remained unsolved for a long time. Yet these changes do not alter the fundamentally recombinant nature of knowledge described above; rather, AI helps scientists navigate the nearly infinite combinatorial space of possible idea combinations and identify promising connections between distant knowledge fields. The assisting role of AI can be clearly seen as papers sometimes show authors crediting LLMs for ``excellent research assistance'' (for example, see \citet{gans2025growth} and \citet{gans2025quest}).

While \citet{griffin2024new} suggest that we may be entering a ``\textit{new golden age of discovery}'' and there is undeniable excitement -- at least from a part of the scientific community -- about AI's potential to foster discovery, there is not a general consensus on its effects \citep{bianchini2026ai}. In particular, the use of AI in research has raised concerns with regard to the loss of skills for scientists over-relying on AI output, the direction of research and the originality of scientific production. These concerns arise because AI systems perform well in data-rich and well-explored domains, potentially directing research efforts toward a restricted set of topics, a phenomenon referred to as \textit{streetlight effect} \citep{tranchero2024b}. Moreover, scientists employing similar tools may converge on similar research directions, increasing the risk of producing duplicate research and reducing novelty, the so-called \textit{stepping-on-toes effect} \citep{almeida2024}.

In the present paper, we refer to \textit{distance} between ideas as their level of conceptual dissimilarity. Ideas can be embedded in a graph where they are represented by nodes and link weights measure their pair-wise distance. We refer to such a graph as \textit{knowledge space}. Pairs of ideas are combined to produce new knowledge and we think to long-distance recombinations as explorative research which combines dissimilar concepts. Conditional on success, they likely lead to more radical innovations. By converse, short-distance recombinations constitute exploitation of already existing ideas and more incremental innovation. We then investigate the following research question: \textit{does the use of AI as a research assistant encourage more radical knowledge recombinations or does it redirect research toward more incremental advances?}

To answer this question, we develop a model of recombinant innovation embedded in a Schumpeterian quality-ladder framework. Innovations arrive according to a Poisson
process, with an individual discovery intensity that decreases with recombinant distance. R\&D firms choose how far to reach across the knowledge space when combining ideas facing a fundamental trade-off: distant recombinations are riskier but, conditional on success, generate more radical and profitable innovations, whereas nearby recombinations are easier to develop but produce more incremental advances. 

Research is modeled through a task-based production function in which a fraction of tasks are AI-automated, while the remaining tasks require human researchers. Each successful innovation grants the innovator a temporary monopoly until it is displaced by the next innovation. AI facilitates recombination at all distances, with a larger proportional effect on the discovery intensity of more distant recombinations.

However, AI also empowers competitors: by raising the aggregate arrival rate of innovations across the economy, it accelerates the rate of creative destruction and it shortens the expected duration of innovation monopolies, thus reducing the expected reward from risky long-distance recombinations. The central contribution of our model is to show how this competitive response shapes recombinant distance: even when AI makes distant recombinations relatively easier, the resulting
replacement threat can lead firms to choose shorter
distances. The direction of research therefore depends
on the balance between improved recombination capabilities
and the endogenous threat of displacement. Also, differently from previous quality ladder models, our framework focuses not on the amount of research carried out in the economy, but on the distance of knowledge recombination, allowing us to analyze whether AI steers research toward more explorative or exploitative research efforts.

Our analysis yields three main results. First, the effect of increasing the share of AI-automated research tasks can be non-monotonic: at low levels of automation, expanding AI usage encourages more explorative recombinations, but at high level of automation, creative destruction dominates and research firms shift their focus toward more exploitative combinations. Second, even with decreasing optimal distance along the balanced growth path, the overall expected log-growth of the knowledge stock may continue to increase thanks to a higher frequency of innovation made possible by the AI; however, our numerical simulation also shows that knowledge growth may slow down if distance falls fast enough. Third, when the knowledge increase following an innovation is below a certain threshold, then the decentralized optimal distance along the balanced path is locally sub-optimal and a planner could instead increase social welfare choosing a greater distance: under the private optimum, research is too incremental.

The remainder of the paper is organized as follows. Section 2 reviews the relevant literature and positions our contribution. Section 3 introduces the model, describing the knowledge space, the task-based R\&D technology, and the role of AI. Section 4 characterizes the stationary equilibrium, and Section 5 derives the comparative statics results. Section 6 presents a numerical simulation of the model. Section 7 provides the welfare analysis, and Section 8 concludes.

\section{Background and Related Literature}
\subsection{Knowledge Creation, Recombination, Exploratory Search and Artificial Intelligence} 
Knowledge recombination is the process through which new knowledge and innovations are created by reconfiguring or combining already existing knowledge and it is a central concept in the innovation literature since \citet{schumpeter1934}. \citet{weitzman1998} builds an analogy with plant cross-pollination with ideas seen as seeds which are combined to create new varieties of knowledge. Newly created concepts are then added to the stock of knowledge and used for future cross-pollinations. As the number of combinations is nearly infinite, the role of researcher is to identify the most promising ones, that is those leading to useful innovations. 

Knowledge creation is an intrinsically risky activity and the uncertainty deriving from the search process is related to the type of combination pursued by researchers. Indeed, \citet{fleming2001recombinant} finds that while researchers can combine familiar components in established ways, introduce unfamiliar components, or connect bodies of knowledge that have rarely been considered together, experimentation with new components and new combinations is on average associated with the production of less useful inventions, but with a greater variability in outcomes, creating opportunities for both failure and breakthrough.

Firms engaging in a search process face a fundamental exploration-exploitation trade-off. Exploitation refers to the refinement of existing knowledge and innovations, whereas exploration concerns the creation of new ones. As the benefits of exploration are more uncertain and postponed in the future, the search process may strengthen exploitation more rapidly than exploration, improving near-term performance while potentially weakening long-run performances, especially in fast-changing environments \citep{march1991}. In a recombinant setting, this trade-off concerns how far search moves beyond an established knowledge base. Local search can exploit accumulated expertise and known relationships between components, while distant search can expose researchers to less familiar possibilities and these activities need not be mutually exclusive. \citet{katilaAhuja2002} show evidence from the robotics industry in which both search scope, defined as exploration of new knowledge and search dept, understood as reuse of existing knowledge are beneficial to the introduction of new products. In scientific research, \citet{uzzi2013atypical} find that high-impact papers tend to blend both familiar and unfamiliar knowledge from prior works. \citet{kaplan2015double} distinguish an invention's cognitive novelty from its subsequent economic value: patents originating new topics are associated with local search, whereas economic value is associated with novelty and broader recombination.

Together, these findings suggest that exploiting established knowledge and introducing unfamiliar elements can be complementary. The model we present captures a distance-based dimension of the exploration-exploitation trade-off: more distant recombinations are harder to obtain but, conditional on success, produce larger knowledge gains and lead to more radical innovations. This is a stylization of such a trade-off, rather than an empirical claim that exploration always has a higher expected payoff.

Over the past decade, artificial intelligence (AI) has been increasingly integrated into scientific activity assisting researchers in processing large datasets, generating hypotheses, designing experiments and accelerating experimentation \citep{wang2023}. Moreover, AI-based predictive models help scientists navigate the vast search space of possible combinations. \citet{agrawalEtAl2019needles} make this clear by showing that advances in algorithms may guide scientists towards knowledge combinations that will yield valuable discoveries.  \citet{agrawal2023, agrawal2024} model innovation as a multi-stage search process and show that improvements in AI may help prioritize candidates with the highest expected returns and reduce both costs and search duration. More specifically, AI-based predictive model can rank candidate molecules, materials or technological designs before costly testing.

AI also provides researchers with knowledge outside its immediate domain of expertise, potentially expanding the range of combinations that can be considered, helps find patterns in complex datasets and Large Language Models (LLMs) have shown potential for theory building and in silico experimentation \citep{tranchero2024b}. In general, AI has been characterized as an ``\textit{invention in the methods of inventing}'', that is a General Purpose Technology (GPT) which is reshaping how research is conducted \citep{venturini2022, bianchini2022, besiroglu2024}. 

Emerging empirical evidence also points to important limitations of AI-assisted scientific research. \citet{liu2023generative} find that although generative AI can address problems in material science, complicated questions involving domain-specific knowledge can not be effectively handled. The need for domain knowledge and interpretability of results still requires the presence of experts in the field. Similarly, in pharmaceutical R\&D, AI accelerates drug discovery but the effect is strongest for drugs with medium levels of novelty and for firms with deep domain knowledge that can filter out false positives \citep{lou2021, tranchero2024a}. More broadly, there is growing concern that AI-assisted research may concentrate on well-explored knowledge domains where data is abundant and AI systems perform best, the so-called streetlight effect \citep{tranchero2024b}. As a consequence, researchers using similar AI tools may converge on the same ideas, producing duplication of research effort, the stepping-on-toes effect \citep{almeida2024}. 

In line with these concerns, \citet{hao2026artificial} use data on scientific papers from 1980 to 2025 to show that while AI adoption has accelerated the rate of discovery in natural sciences and AI-adopters publish 3.02 times more papers than non-adopters, the volume of studied topics has narrowed by 4.63\%. Evidence from specific research settings suggests that AI can also redirect attention toward less-explored areas. Studying the introduction of AlphaFold2, \citet{hill2026artificial} find that downstream basic research increased on proteins for which structural information had previously been unavailable, relative to proteins with existing structural information. \citet{gans2025growth, gans2025quest} model AI as a technology which interpolates between existing knowledge points and study whether it pushes more explanatory or incremental research. In both papers, he finds that the effect of AI is non-monotonic and depends on its interpolating capabilities. When AI power is limited or moderate, it reduces the cost of working within well-understood domains, so scientists favor incremental research that deepens existing knowledge; only when AI is sufficiently capable, scientists purue more novel research. This capability threshold also drives growth: modest AI raises growth when ideas are getting harder to find, whereas highly capable AI promoting exploratory research lowers knowledge density and can slow growth despite direct productivity gains. 

Taken together, these findings suggest that AI affects not only scientists' productivity, but also the general direction of scientific research and that the relationship between AI adoption and research novelty may be more complex than simple productivity-enhancing accounts would imply.

\subsection{AI, Productivity and Economic Growth}
Most of the literature on automation focuses on the effects on production, productivity and growth. \citet{acemogluRestrepo2019} show that automation displaces workers from tasks they previously performed, while also generating positive productivity effects, so that its net impact on labor demand depends on which force prevails. In addition, the creation of new tasks in which human workers retain a comparative advantage reinstates labor demand. As a result, productivity gains need not translate into proportional changes in employment or in factor income shares. \citet{acemoglu2025simple} estimates the macroeconomic effects of generative AI by combining evidence on task exposure with estimates of cost savings, and finds modest gains in TFP and GDP over the next ten years. This follows from the fact that only a limited share of tasks can be profitably automated, and that the available evidence on cost savings comes mostly from tasks that are relatively easy for AI, whereas for more complex and context-dependent tasks, AI implementation will be more difficult.

A growing body of literature is also studying the effect of AI on research activity. \citet{aghion2017} examine automation in both production and research and finds that although AI could replace people in generating ideas and in principle could lead to a growth singularity, in fact growth remains bounded by the sectors which are difficult to improve. \citet{jones2025} also emphasizes Baumol's cost disease: if some steps of the research process are hard to improve, they become bottlenecks that limit the impact of AI on idea production. \citet{jonesTonetti2026} propose an endogenous growth model where both production of goods and of ideas can be AI-automated but an elasticity of substitution among tasks lower than one slows economic growth due to slowing-improving labor. \citet{jones2026future} further emphasizes the role of ``weak links'' in preventing growth acceleration that one could expect following AI-automation of goods and idea production. In contrast with these works, \citet{davidson2026automating} develop a semi-endogenous growth model with an innovation network in which a growth singularity is possible even in the presence of bottlenecks, provided that task automation advances sufficiently fast.

\subsection{Creative Destruction and Position of the Present Model}
\citet{gans2025growth} assigns research firms exclusive rights to develop their respective frontiers and studies competition in AI provision, while \citet{gans2025quest} introduces rivalry among scientists as an extension. \citet{cockburn2018impact} discuss AI as a general-purpose method of invention and highlight incentives for firms to race to acquire and control critical datasets and application-specific algorithms. Their analysis suggests that AI may reshape not only research productivity but also the organization of R\&D and the competitive environment. 

However, the role of competition among innovators in shaping AI-induced changes in research direction remains underexplored. Our work places this competition at the centre of a quality-ladder framework, examining how AI interacts with endogenous research entry and the threat of displacement to shape recombination choices and economic growth. More specifically, we combine the distance-based search question with the creative-destruction mechanism of quality-ladder models \citep{aghionhowitt1992,aghionhowitt1998}. In this framework, AI empowers research firms to attempt more explorative recombinations
and successful innovators obtain temporary monopoly rents, until displaced by the next innovator. Our model makes recombinant distance central to this incentive problem as a more distant success generates a larger knowledge gain and monopoly reward, but also creates more productive opportunities for subsequent researchers. While on one side AI increases technological feasibility of long-distance recombinations, it also increases the overall arrival rate of innovations, introducing a tension between long and short distance research, analogously to the exploitation-exploration trade-off.

We employ a task-based idea production function as in \citet{jones2025}, where an exogenous fraction of tasks is AI-automated and we study how increasing this fraction changes research costs and the relative technological feasibility of distant recombination, while free-entry determines the mass of active research units.


\section{The Quality-Ladder Model}
\subsection{The Knowledge Space}
\begin{figure}
    \centering
    \begin{subfigure}{\linewidth}
        \centering
        \includegraphics[width=0.7\linewidth]{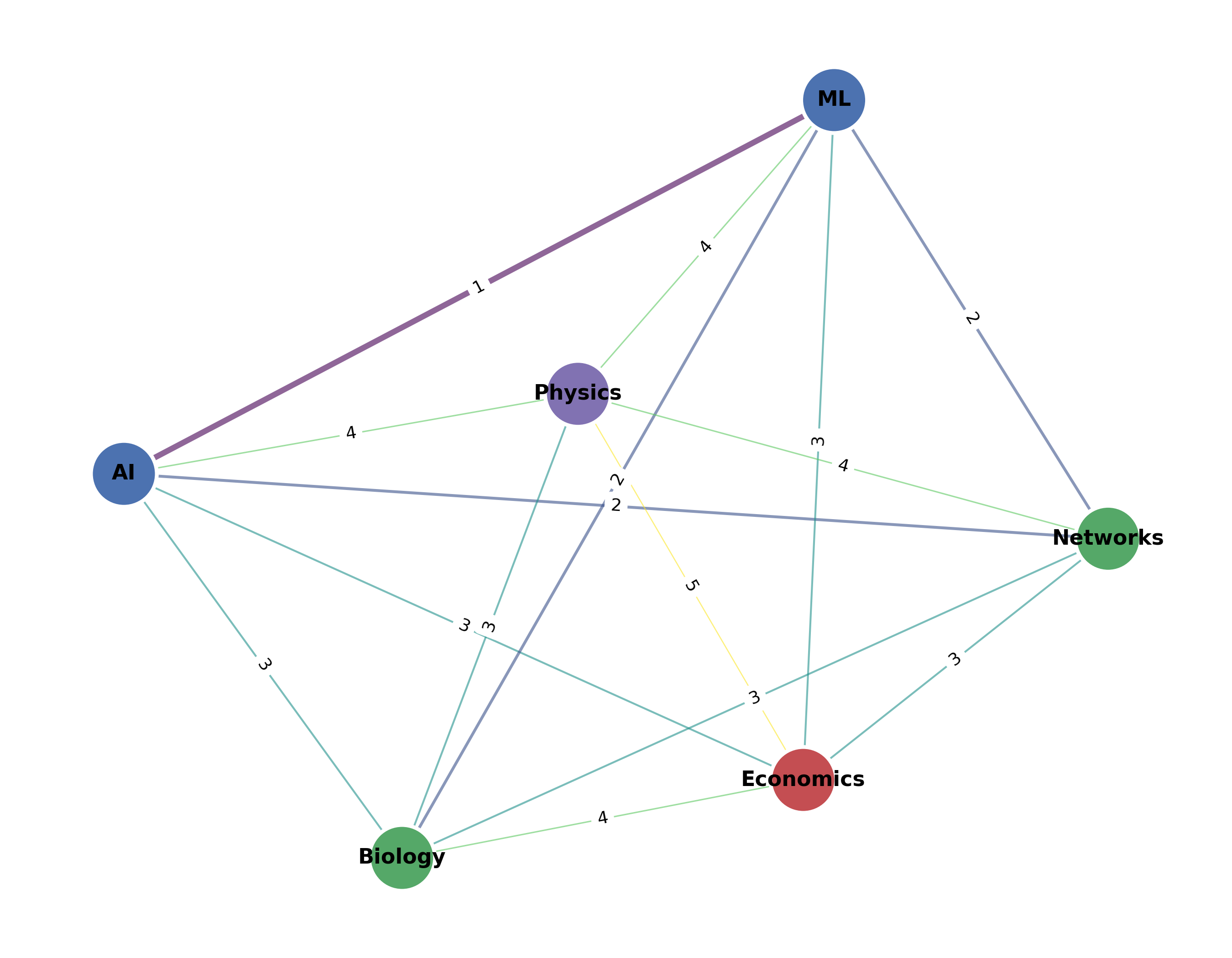}
        \caption{Knowledge space}
        \label{a}
    \end{subfigure}
    \hfill
    \begin{subfigure}{\linewidth}
        \centering
        \includegraphics[width=0.7\linewidth]{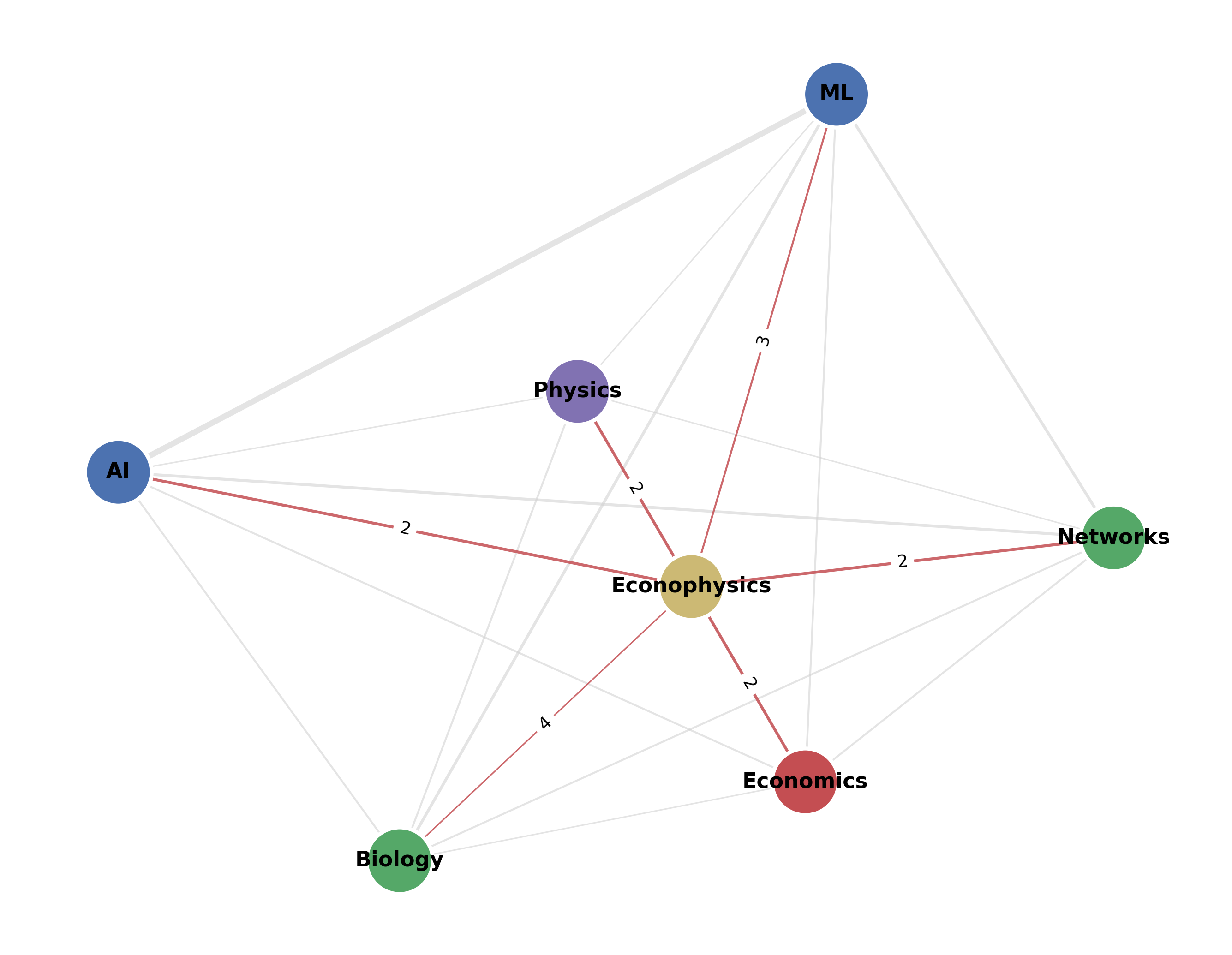}
        \caption{Knowledge space with a new bridging field}
        \label{b}
    \end{subfigure}
    \caption{Knowledge space before (above) and after (below) the introduction of the new ``Econophysics'' node. Link weights indicates the distance among knowledge fields; red links are the new combination possibilities introduced by the new field.}
    \label{fig:knowledge_space}
\end{figure}

Let $\mathcal{G} = (\mathcal{I}, \mathcal{D})$ define \textit{knowledge space}, that is the graph of existing ideas, where $\mathcal{I}$ denotes the set of nodes, each corresponding to an individual idea or ,for simplicity, to a research field, and $\mathcal{D} = \{ d_{ij} \}_{i,j \in \mathcal{I}}$ is the set of pairwise distances between them. Distances capture the degree of cognitive dissimilarity between ideas: a large distance means that two ideas pertains to very different knowledge domains and they are therefore also more difficult to combine. Knowledge similarity can be measured, for instance, using patent citation data \citep{chen2017} or through text-based analyses of scientific publications that capture semantic similarity across research domains \citep{gentzkow2019}. Fig. \ref{a} shows an example of knowledge space. 

\subsection{R\&D Sector}
R\&D firms produce new knowledge by combining knowledge and ideas from different fields (recombinant innovation). To do so, they use both human researchers and AI. AI helps researchers to identify complex patterns, perform routine and repetitive cognitive tasks more efficiently, and provide access to advanced scientific knowledge (and particularly that from distant scientific fields outside of the firm's core domain of expertise). 

To attempt a knowledge recombination, the firm must perform a set of tasks $\omega \in [0,1]$.  Let us denote $y(\omega)$ the generic output from a given task, and let us define the total research effort $R$ as:
\begin{equation}
R = \exp \left( \int_0^1 \log y(\omega)\, d\omega \right), 
\label{ipf_1}
\end{equation}
with
\begin{equation}
y(\omega) =
\begin{cases}
m \frac{X_{\text{AI}}}{\alpha} & \text{for $\omega \in [0, \alpha] $} \\
\frac{l_R}{1-\alpha} & \text{for $\omega \in (\alpha, 1]$}
\end{cases},
\end{equation}
where an exogenously given fraction $\alpha\in(0,1)$ of tasks is performed by AI, while the remaining fraction $(1-\alpha)$ must be performed using labor; $m$ is AI productivity and, $x_{\text{AI}}$ and $l_R$ are the total AI power and the total number of researchers, respectively, demanded by a given firm.

Substituting these inputs into eq. \eqref{ipf_1} gives:
\begin{equation}
    R = \frac{(m X_{\text{AI}})^\alpha l_R^{(1-\alpha)}}{\alpha^\alpha(1-\alpha)^{(1-\alpha)}}.
\end{equation}

Our focus is to study the optimal recombinant distance, not the intensive margin of research. Therefore, for simplicity, we fix total research effort $R = \bar{R} =1$. Hence, $\bar{R}$ represents the effort required to attempt a single recombination.

The firm's optimization problem is to minimize its input costs to achieve a recombinant innovation, taking wage $w$, AI price $\mu$ and the knowledge space as given:
\begin{equation}
\begin{aligned}
\min_{l_{R,t}, X_{\text{AI},t}} \quad & w_tl_{R,t} + \mu_t X_{\text{AI},t} \\
\text{s.t.} \quad & \bar{R} = 1. 
\end{aligned}
\label{research_min_prob}
\end{equation}
Solving this problem, we get the following expression for the recombinant cost:
\begin{equation}
C(w_t, \mu_t) = \frac{w_t^{1-\alpha} \mu_t^\alpha}{m^{\alpha}}.
\label{cost_function}
\end{equation}

Following \citet{jones2025}, we define what he calls ``technology adoption condition'':
\begin{equation}
\frac{\mu}{m} \le w, \quad \forall\omega\in(0, \alpha).
\label{tech_adop_cond}
\end{equation}
This condition ensures the economic convenience on using AI on automatable tasks. For the baseline analysis, we also assume that AI price scale proportionally with knowledge stock. Together with the assumption that researchers represent a small fraction of the total labor supply, this keeps the AI-to-labor price ratio constant and supports a Balanced Growth Path with constant optimal distance. We discuss possible alternative pricing solution in Appendix \ref{app:ai_pricing}.

\paragraph{\textbf{Assumption 1} (AI pricing)}
AI price $\mu>0$ grows proportionally with the stock of knowledge $A$, that is $\mu \propto A$.

A possible supply-side justification is that AI is produced competitively using labor with constant productivity $a_{AI}$.  Competitive AI pricing gives $\mu(A) = w(A)/a_{AI}$ and by assuming that labor employed in research and AI production is small relative to final good production allows us to neglect their feedback on the determination of equilibrium wage.

\subsection{AI, Discover Difficulty and Opportunities, and Innovation Quality}
Let define $p(d; \alpha, m)>0$ the base discovery intensity, i.e. the base Poisson arrival rate for a recombination at distance $d$, capturing a decreasing innovation intensity as recombinant distance increases. Let us define:
\begin{equation}
    s(d; \alpha, m) = - \frac{\partial \ln p(d; \alpha, m)}{\partial d} = -\frac{p_d}{p} >0,
    \label{s}
\end{equation}
measuring the discovery-intensity penalty associated with a marginal increase in recombinant distance.

We claim that nearby ideas are easy to recombine and AI facilitates recombination at all distances, with a larger proportional effect on the discovery intensity of more distant recombinations, thereby reducing their relative technological disadvantage:
\begin{equation}
p_\alpha \ge 0, \quad s_\alpha < 0.
\end{equation}

Let $A_t>0$ denote the stock of knowledge in the economy at period $t$, where $A_t$ encompasses the whole amount of ideas and information accumulated up to that period. We also define a period as the amount of time elapsing between two consecutive innovations as in \citet{aghionhowitt1992, aghionhowitt1998}, so that at period $t$, the $t$-th innovation as been introduced into the economy. Each time a recombination is successful, the stock of knowledge is improved by a factor $\gamma(d)\ge0$, capturing the higher number of ideas and information in the economy and therefore, the productivity increases in the production sector. In particular, we claim more distant recombination, although being less frequent are more valuable as they introduce substantial new fields or line of research. Thus, we have:
\begin{equation}
\begin{aligned}
    &A_{t+1} = AQ(d_{t}), \\
    &\text{with} \quad Q(d_{t})=1+G(d_t) \quad \text{and} \quad G(d) = \gamma(d).
    \label{knowledge_increase}
\end{aligned}
\end{equation}

We also define the marginal proportional quality gain of knowledge as:
\begin{equation}
    h(d)= \frac{\partial\ln Q(d)}{\partial d} = \frac{\gamma'(d)}{Q(d)}.
    \label{h}
\end{equation}

Every time a research firm successfully recombines knowledge and introduces an innovation, it contributes on the total stock of knowledge and gets a monopoly on an upgraded version of the (unique) intermediate good used in final good production. The monopoly lasts until displaced by the next innovator and at that point, the previous innovator will not get any profit from its innovation.

Finally, a current innovation opens up new opportunities for further discoveries. These research-opportunity spillovers are captured by $\psi(d)$, with $\psi'(d)\ge0$. We define the proportional marginal increase in opportunities following a marginal increase in latest recombinant distance as:
\begin{equation}
      u(d) = \frac{\partial \ln\psi(d)}{\partial d} = \frac{\psi'(d)}{\psi(d)}.
      \label{u}
\end{equation}

Although building a dynamic network model of knowledge recombination is outside the scope of the present paper, we can nevertheless build the following useful analogy.  Consider the birth of ``Econophysics'', the inter-disciplinary field where theories and methods from physics have been applied to study economic problems with particular attention to financial markets \citep{sharma2011econophysicsbriefreviewhistorical}, shown in Fig. \ref{b}. The success of a recombination opens new paths that may lead to further combinations in the future, in line with \citet{weitzman1998}'s idea. In particular, we claim that the combination between seemingly distant fields may open up much more opportunities for subsequent research. The opening of new recombinant paths from the latest innovation is summarized, in a reduced form, by the opportunity spillovers factor $\psi(d)$. For tractability, opportunity spillovers from previous recombinations do not compound, but are embodied in the stock of knowledge $A_t$. This mechanism is similar to the knowledge spillovers in \citet{liu2021innovation}, where one sector's research activity builds on prior discoveries, often coming from other scientific fields. In our framework, we assume that the strength of this effect increases with recombination distance. 

Hereafter, we present the hypotheses supporting our model:

\paragraph{\textbf{Assumption 2} (Arrival of innovation)} $p(d; \alpha, m)>0$ and smooth on positive distances, $s(d; \alpha, m) >0$ and $s_d(d; \alpha, m)\ge 0$. Further, $\lim_{d\rightarrow0^+}p(d; \alpha, m)=p_0 \in (0, \infty)$ and $s(d; \alpha,m)$ has a finite limit. For sufficiently large $d$, $s(d)\ge\underline{s}>0$.

\paragraph{\textbf{Assumption 3} (Quality)} $\gamma(d)$ is continuous on $[0, \infty)$ and $C^2$ on $(0, \infty)$, with $\gamma(0)=0$, $\gamma'(d)>0$, $\gamma''(d)\le0$ and $\lim_{d\rightarrow0^+} \gamma'(d) = \infty$.

\paragraph{\textbf{Assumption 4} (Opportunities)} $\psi(d)>0$ and $C^2$, with $\lim_{d\rightarrow0^+}\psi(d) \in (0, \infty)$, $u(d)\ge0$, $u_d(d)\ge0$ and $\lim_{d\rightarrow0^+}u(d)$ finite.

\subsection{Final Good Sector}
Following \citet{aghionhowitt1992, aghionhowitt1998}, there is a single final good produced according to the following production function:
\begin{equation}
    Y_t = A_t y_t^\epsilon,
    \label{fin_good_prod}
\end{equation}
where $\epsilon \in (0, 1)$ and $y_t$ is the amount of (updated) intermediate good that final goods firms purchase from R\&D companies following their successful innovation. Since the final good market is competitive, final good producers take the intermediate good price as given and demand it up to the point where its price equals its marginal product. This yields the inverse demand function: $p_t = \epsilon A_t y_t^{\epsilon-1}$.

The latest innovator who becomes monopolist of the intermediate good determines the optimal quantity to supply by maximizing its profit: $\pi_t = p_t y_t - y_tw_t$, giving:
\begin{equation}
y_t^* = \left( \frac{\epsilon^2A_t }{w_t} \right)^{\frac{1}{1-\epsilon}} \quad \text{and} \quad \pi^* = \epsilon(1-\epsilon)Ay^\epsilon.
\label{optimal_quantity}
\end{equation}

Assuming for tractability that researchers (and workers employed in AI production) are a negligible fraction of the total labor supply in the economy ($\bar{L}$), i.e. $L_R\ll \bar{L}$, we get:
\begin{equation}
    w = \iota A \quad \text{and} \quad \pi = \Pi A 
\end{equation}
where $\iota \equiv \epsilon^2 \bar{L}^{\epsilon-1}$ and $\Pi \equiv \epsilon(1-\epsilon)\bar{L}^\epsilon$.

Under Assumption 1, we can express AI price as $\mu_t = A_t \bar{\mu}$ and we can rewrite the equation of research cost \eqref{cost_function} as:
\begin{equation}
    C(w_t, \mu_t) = A_t \iota^{1-\alpha} (\bar{\mu}/m)^\alpha \equiv A_t\Gamma.
\end{equation}

\subsection{Poisson Arrivals}
Let $N(z)$ the mass of active research firms in a given period, after last innovation's distance $z$. We introduce congestion in research activity summarized by an increasing function $\xi(N(z))$, capturing the fact that more research units decrease each unit's individual discovery rate.

Any research firm $i$ now targeting a combination at distance $x$ has a Poisson arrival rate of successful recombinations defined by:
\begin{equation}
    \rho_i(x \mid z)
    =
    \frac{\psi(z)}{\xi(N(z))}p(x; \alpha, m).
\end{equation}

If all firms target the same distance $x$, then the aggregate Poisson arrival of successful recombination in the economy is:
\begin{equation}
    H(x \mid z)
    =
    \frac{N(z)\psi(z)}{\xi(N(z))}p(x; \alpha, m).
    \label{aggregate_poisson}
\end{equation}

For subsequent analysis, we define the congestion elasticity as:
\begin{equation}
    e(N)
    =
    \frac{N\xi'(N)}{\xi(N)},
    \label{e}
\end{equation}
and we state the following assumption:

\paragraph{\textbf{Assumption 5} (Congestion)}
$\xi(N)>0$ and $C^2$ on $(0, \infty)$, $\xi'(N)>0$.
Further, $\lim_{N \rightarrow 0^+}\xi(N)=0$ and
$\lim_{N \rightarrow \infty}\xi(N)=\infty$.
Finally, $0<e(N)<1$ and $e'(N)\leq 0$.

\section{Private Decisions and Equilibrium}

\subsection{Distance Optimization}

Let $z>0$ denote the distance of the last innovation (the incumbent monopolist) and let $b(z)$ denote the distance targeted by researchers operating in that inherited state. An individual research firm chooses its own
distance, while it does not chooses the future mass of research firms, nor their target distance in case of its own success.

The incumbent monopolist earns intermediate-good profit
$\Pi A$ until replaced and zero thereafter because it does not receives royalties for previous innovations. Its monopoly
value is therefore
\begin{equation}
    V(A,z)
    =
    \frac{\Pi A}{r+H(b(z)\mid z)},
    \label{incumbent_value}
\end{equation}
where $r>0$ is the exogenous discount rate.
The replacement intensity depends on the target $b(z)$
of current researchers and the inherited state $z$.

A current research firm chooses a distance $x>0$.
Conditional on success, knowledge becomes $AQ(x)$ and
the inherited state becomes $x$. The mass of subsequent researchers is given by $N(x)$, each targeting distance $b(x)$. 
Any research firm's expected net payoff rate is
\begin{equation}
    J(x\mid A,z)
    =
    \frac{\psi(z)}{\xi(N(z))}
    \frac{p(x)\Pi A Q(x)}{r+H(b(x)\mid x)}
    -A\Gamma,
    \label{research_return}
\end{equation}

We suppress the dependence of $p$ on the exogenous
parameters $\alpha,m$, which are held fixed.

The research firm solves
\begin{equation}
    \max_{x>0}J(x\mid A,z).
    \label{max_problem}
\end{equation}

In eq. \eqref{research_return}, we see that the positive factor $A\psi(z)/\xi(N(z))$ multiplies
returns at every candidate distance equally, and
$A\Gamma$ is independent of $x$. Consequently, the set of privately optimal distances is the same across current inherited states. If the maximizer is unique, a consistent research-target
rule must therefore be constant across those states. Thus, $b(z)=\bar{d}$.

Solving problem \ref{max_problem} is equivalent to maximize the logarithm of the gross-return:
\begin{equation}
    L(x)
    =
    \log p(x)+\log Q(x)
    -\log\bigl(r+H(\bar d\mid x)\bigr).
    \label{private_log_return}
\end{equation}

An interior solution must satisfy the following first-order condition:
\begin{equation}
    h(x)-s(x)
    -
    \frac{1}{r+H(\bar d\mid x)}
    \frac{\mathrm d}{\mathrm d x}H(\bar d\mid x)
    =0.
    \label{distance_foc}
\end{equation}

The anticipated target distance $\bar d$ chosen by subsequent researchers remains constant, while future entry $N(x)$ responds to the state created by current success. The three terms in \eqref{distance_foc} represent the proportional quality gain, the proportional discovery penalty, and the marginal replacement effect on monopoly value, respectively.

\subsection{Balanced Growth Path Equilibrium}
A symmetric equilibrium consists of a common optimal
target distance $d^*$ and a mass of active research
firms $N(z)$ for every admissible inherited state $z$.
Research firms can freely enter or exit without
adjustment costs.

The target anticipated for subsequent researchers
must coincide with the distance firms optimally
choose: $\bar d=d^*$. Private optimality and free
entry with positive research activity require,
respectively,
\begin{equation}
    d^*\in
    \operatorname*{arg\,max}_{x>0}J(x\mid A,z),
    \qquad
    J(d^*\mid A,z)=0,
    \label{equilibrium_optimality}
\end{equation}
where $J$ is evaluated with subsequent researchers
targeting $d^*$. Evaluating the researcher's choice
at $x=d^*$ and imposing zero profit yields
\begin{equation}
    \frac{\psi(z)}{\xi(N(z))}
    p(d^*)
    \frac{\Pi Q(d^*)}{r+H(d^*\mid d^*)}
    =
    \Gamma.
    \label{free_entry}
\end{equation}

Rearranging \eqref{free_entry} yields:
\begin{equation}
        \frac{\psi(z)}{\xi(N(z))}
        =
        \frac{\Gamma\bigl(r+H(d^*\mid d^*)\bigr)}
             {p(d^*)\Pi Q(d^*)}
        \equiv k,\\
    \label{continuation}
\end{equation}
Within one equilibrium, $k$ is constant for any inherited state. Combining \eqref{aggregate_poisson} and \eqref{continuation} allows us to express the aggregate arrival rate as:
\begin{equation}
    H(d^*\mid z)=kp(d^*)N(z).
    \label{continuation_intensity}
\end{equation}

Although the inherited state $z$ does not influence the current optimal choice of recombination distance, it influences the replacement rate through the opportunities open up by the latest innovation (via $\psi(z)$) and the mass of active research firms (via $N(z)$). A researcher considering a candidate distance $x$ anticipates these effects on subsequent research and, consequently, on its own monopoly duration. Conditional on success at $x$, subsequent researchers inherit state $x$, operate in mass $N(x)$, and target $d^*$.
We can see it differentiating eq. \eqref{continuation_intensity} with respect to the inherited state $z$:
\begin{equation}
    \frac{1}{H(d^*\mid z)}
    \frac{\mathrm d}{\mathrm d z}H(d^*\mid z)
    =
    \frac{N'(z)}{N(z)}
    =
    \frac{u(z)}{e(N(z))}.
    \label{eq:continuation_response}
\end{equation}

See Appendix~\ref{subsec:replacement_change}
for the derivation.

Along the BGP, the inherited distance equals the common current (and subsequent) target $d^*$. Define
\begin{equation}
\begin{aligned}
    &N^*=N(d^*),\\
    &H^*=H(d^*\mid d^*)=\frac{N^*\psi(d^*)}{\xi(N^*)}p(d^*).
    \label{bgp_values}
    \end{aligned}
\end{equation}

Private optimality condition and free-entry give:
\begin{align}
    h(d^*)-s(d^*)
    -
    \frac{u(d^*)}{e(N^*)}
    \frac{H^*}{r+H^*}
    &=0,
    \label{equilibrium_distance}\\
    \frac{\psi(d^*)}{\xi(N^*)}
    p(d^*)
    \frac{\Pi Q(d^*)}{r+H^*}
    &=\Gamma.
    \label{stationary_free_entry}
\end{align}

Equations \eqref{equilibrium_distance} and
\eqref{stationary_free_entry} jointly determine the
stationary pair $(d^*,N^*)$ and we present the following propositions:

\paragraph{\textbf{Proposition 1} (Unique private optimal distance)}
Suppose Assumptions 1--5 hold. For given finite $k>0$ and subsequent target $\bar d>0$, let subsequent research activity satisfy $N(z)=\xi^{-1}(\psi(z)/k)$ and $H(\bar d\mid z)=kp(\bar d)N(z)$. The private first-order condition
\eqref{distance_foc} has a unique solution
$x^*>0$. This solution is the unique interior
global maximizer of the researcher's expected net payoff rate
\eqref{research_return}.

The proof is provided in Appendix~\ref{proof:private}.

\paragraph{\textbf{Proposition 2}
(Existence and uniqueness of the stationary equilibrium)}

Under Assumptions 1--5, there exists a unique
positive stationary equilibrium pair $(d^*,N^*)$ defined by equations \eqref{equilibrium_distance} and
\eqref{stationary_free_entry}, with aggregate
discovery intensity given by \eqref{bgp_values}.

The proof is provided in Appendix~\ref{proof:existence}.

\subsection{Balanced Growth Path and Knowledge Growth}
In event time, where period $t$ denotes the arrival of the $t$-th innovation,
knowledge grows by a factor $Q^*$ between consecutive periods along the
balanced growth path, i.e., $A_t = Q^*A_{t-1}$. To express growth in calendar
time, let $\nu$ denote time measured in years and let $M(\nu)$ be the number of
innovations that have arrived by $\nu$, with $M(0)=0$. Since innovations arrive
according to a Poisson process with rate $H^*$ per year,
\begin{equation}
  A(\nu) = A_0\,(Q^*)^{M(\nu)}, \qquad M(\nu+1)-M(\nu)\sim \text{Poisson}(H^*).
\end{equation}
Calendar-time growth is therefore random, since it depends on the number of
successful recombinations in a year. Taking logs,
\begin{equation}
  \ln A(\nu+1)-\ln A(\nu) = \big[M(\nu+1)-M(\nu)\big]\ln Q^*,
\end{equation}
so the expected annual log-growth is
\begin{equation}
  g^*_{\ln A} \equiv \mathbb{E}\big[\ln A(\nu+1)-\ln A(\nu)\big] = H^*\ln Q^*.
\end{equation}

\section{Comparative Statics}
We now study the effect of an increase in the exogenous fraction of automated tasks $\alpha$ on the equilibrium distance, research mass and knowledge growth rate. We define the following functions (also defined in Appendix \ref{proof:existence}) making the dependence on parameter $\alpha$ explicit:
\begin{equation}
\begin{aligned}
    \rho(d,N;\alpha)
    &:=
    \frac{\psi(d)}{\xi(N)}p(d;\alpha,m), \qquad
    \mathcal H(d,N;\alpha):=
    N\rho(d,N;\alpha), \qquad
    S(d;\alpha)
    :=
    \frac{\Pi Q(d)}{\Gamma(\alpha,m)}, \\
    \tau(d,N)
    &:=
    \frac{u(d)}{e(N)}, \qquad
    \chi(d,N;\alpha)
    :=
    \frac{\mathcal H(d,N;\alpha)}
         {r+\mathcal H(d,N;\alpha)},
\end{aligned}
    \label{cs_definitions}
\end{equation}
and we rewrite the distance condition \eqref{equilibrium_distance} and free-entry \eqref{stationary_free_entry} as:
\begin{align}
    F(d,N;\alpha)
    &:=
    h(d)-s(d;\alpha,m)-\tau(d,N)\chi(d,N;\alpha)
    =0,
    \label{cs_F}\\
    E(d,N;\alpha)
    &:=
    \rho(d,N;\alpha)[S(d;\alpha)-N]-r
    =0.
    \label{cs_E}
\end{align}

We additionally define the direct proportional effects of
automation on discovery intensity and research
cost as:
\begin{equation}
\begin{aligned}
    \ell_\alpha(d;\alpha,m)
    &:=
    \frac{\partial\log p(d;\alpha,m)}
         {\partial\alpha},\\
    c_\alpha
    &:=
    \frac{\partial\log\Gamma(\alpha,m)}
         {\partial\alpha}
    =
    \log\left(\frac{\bar\mu}{m\iota}\right).
\end{aligned}
    \label{cs_direct_effects}
\end{equation}
Given our assumptions, $\ell_\alpha\geq0$, $s_\alpha\leq0$, and $c_\alpha\leq0$, that is automation directly (weakly) increases discovery intensity at a given distance, proportionally advantages more distant recombinations and lowers research costs. We evaluate the partial derivatives of the distance and free-entry conditions and define at $(d^*,N^*;\alpha)$:
\begin{equation}
    \mathcal A:=-F_d>0,
    \qquad
    \mathcal B:=-F_N\geq0,
    \qquad
    \mathcal C:=E_d>0,
    \qquad
    \mathcal D:=-E_N>0,
    \label{cs_coefficients}
\end{equation}

where their signs have been established in Appendix \ref{proof:existence}. Further, at equilibrium, the direct automation effects on the distance \eqref{cs_F} and free-entry \eqref{cs_E} are:
\begin{align}
    F_\alpha
    &=
    -s_\alpha
    -\tau\chi(1-\chi)\ell_\alpha,
    \label{cs_Falpha}\\
    E_\alpha
    &=
    r\ell_\alpha-(r+H^*)c_\alpha.
    \label{cs_Ealpha}
\end{align}
All functions on the right-hand sides are evaluated at the stationary equilibrium. We now state the following proposition to study the effect of a change in $\alpha$ on the equilibrium distance and research firms.

\paragraph{\textbf{Proposition 3}
(Comparative statics of automation)}
Under Assumptions 1--5, the stationary equilibrium is locally continuously differentiable in $\alpha$ at every interior automation level $\alpha\in(0,1)$ and we have:
\begin{align}
    \frac{\mathrm d d^*}{\mathrm d\alpha}
    &=
    \frac{\mathcal D F_\alpha
          -\mathcal B E_\alpha}{\Delta},
    \label{cs_distance_response}\\
    \frac{\mathrm d N^*}{\mathrm d\alpha}
    &=
    \frac{\mathcal C F_\alpha
          +\mathcal A E_\alpha}{\Delta},
    \label{cs_mass_response}
\end{align}
where $\Delta:=\mathcal A\mathcal D+\mathcal B\mathcal C>0.$ is the determinant of the Jacobian matrix build from the equilibrium conditions. The proof is provided in Appendix~\ref{proof:comparative_statics}.\\

The distance response can be decomposed as
\begin{equation}
\begin{aligned}
    \frac{\mathrm d d^*}{\mathrm d\alpha}
    =
    \frac{1}{\Delta}\Bigl\{
    \mathcal D(-s_\alpha)
    -\mathcal D\tau\chi(1-\chi)\ell_\alpha
    -\mathcal B
    \bigl[r\ell_\alpha-(r+H^*)c_\alpha\bigr]
    \Bigr\}.
\end{aligned}
    \label{cs_distance_channels}
\end{equation}

The first term inside the bracket of eq. \eqref{cs_distance_channels} represents the improvement in the relative easiness of combining distant ideas due to greater automation; the second term captures the effect of changing replacement rate at a given distance and research firm mass, and the third term is the change in free-entry condition as greater automation increases discovery rate and lowers research costs therefore giving incentive to enter the research sector.

Consequently, even though greater task automation improves research technology and makes more distant projects relatively easier, equilibrium distance may decrease. Such a decrease occurs if and only if:
\begin{equation}
    \mathcal D(-s_\alpha)<{}\mathcal D\tau\chi(1-\chi)\ell_\alpha
    +
    \mathcal B
    \bigl[r\ell_\alpha-(r+H^*)c_\alpha\bigr].
    \label{cs_distance_decline}
\end{equation}
If the distance response changes sign from positive to negative ad $\alpha$ increases, we may also have a hump-shaped behavior of equilibrium distance, first increasing with the share of automated tasks and then decreasing with it.\\

If research opportunities are independent of distance, so that $u=0$, then $\tau=0$ and $B=0$. Consequently, 
\begin{equation}
    \frac{\mathrm d d^*}{\mathrm d\alpha}
    =
    \frac{-s_\alpha}{s_d-h_d}
    \geq0,
    \label{cs_no_opportunities}
\end{equation}
that is if the research opportunities opened up were  independent on last innovation's distance, then the equilibrium distance would always increase with automation. 

The total response of the aggregate discovery intensity follows from its
technology:
\begin{equation}
    \frac{1}{H^*}\frac{\mathrm d H^*}{\mathrm d\alpha}
    = (u-s)\frac{\mathrm d d^*}{\mathrm d\alpha}
    + \frac{1-e}{N^*}\frac{\mathrm d N^*}{\mathrm d\alpha}
    + \ell_\alpha .
    \label{cs_intensity_response}
\end{equation}

In calendar time, the expected annual log-growth $g^*_{\ln A}=H^*\ln Q^*$
responds to changes in $\alpha$ according to
\begin{equation}
    \frac{\mathrm d g^*_{\ln A}}{\mathrm d\alpha}
    = \ln Q^*\,\frac{\mathrm d H^*}{\mathrm d\alpha}
    + H^*h^*\,\frac{\mathrm d d^*}{\mathrm d\alpha}.
    \label{cs_log_growth}
\end{equation}
Dividing by $g^*_{\ln A}$ and using \eqref{cs_intensity_response}, we get:
\begin{equation}
    \frac{1}{g^*_{\ln A}}\frac{\mathrm d g^*_{\ln A}}{\mathrm d\alpha}
    = \left(u-s+\frac{h^*}{\ln Q^*}\right)\frac{\mathrm d d^*}{\mathrm d\alpha}
    + \frac{1-e}{N^*}\frac{\mathrm d N^*}{\mathrm d\alpha}
    + \ell_\alpha .
    \label{cs_calendar_growth}
\end{equation}

Since at equilibrium $h-s=\tau\chi$, the coefficient on the distance response can be written as
\begin{equation}
    u-s+\frac{h^*}{\ln Q^*}
    = u+\tau\chi+h^*\left(\frac{1}{\ln Q^*}-1\right).
    \label{cs_growth_distance_coefficient}
\end{equation}

The coefficient on the distance response is strictly
positive whenever $\ln Q^*<1$, equivalently
$\gamma(d^*)<\exp(1)-1$.
At $\ln Q^*=1$, it equals $u+\tau\chi$ and is
nonnegative, with strict positivity if
$u+\tau\chi>0$.
Another sufficient condition for strict positivity,
valid for any $Q^*>1$, is $u+\tau\chi\geq h^*$.

Whenever this coefficient is positive, the distance
response contributes to knowledge growth with the
same sign as $\mathrm d d^*/\mathrm d\alpha$.
Thus, a decline in equilibrium distance contributes
negatively to the growth response.
Nevertheless, total growth can increase if the
combined contribution of research-mass adjustment
and the direct technological effect of automation
is sufficiently positive:
\begin{equation}
    \frac{1-e}{N^*}
    \frac{\mathrm d N^*}{\mathrm d\alpha}
    +\ell_\alpha
    >
    -
    \left(u-s+\frac{h^*}{\ln Q^*}\right)
    \frac{\mathrm d d^*}{\mathrm d\alpha}.
    \label{cs_growth_positive_condition}
\end{equation}

\section{Numerical Simulation}
To study the effect of increasing AI automation in, we now show a numerical simulation of the presented model research with the following functional forms:
\begin{equation}
    Q(d) = 1+d^\eta, \qquad \psi(z) = \exp(\theta z), \qquad \xi(N)=N^\delta, \qquad
    p(d; m, \alpha) = p_0 \exp \left(\frac{-\beta d}{\lambda_0+\lambda\alpha}\right),
    \label{functional_forms}
\end{equation}
with $\eta, \delta \in (0,1)$ and $\theta\ge0$. Normalized research cost is given by $\Gamma(\alpha)=\iota c^\alpha$, where $c=\bar\mu/(m\iota)$ is the AI cost relative to labor cost. Table \ref{tab:simulation_parameters} describes the various parameters and the value used in the baseline simulation. Appendix \ref{robust_app} shows the robustness analysis for different parameters values.

Notice that in this specification, greater automation monotonically increases AI effectiveness in recombining ideas in the knowledge space. Any decrease in equilibrium distance is not implied by a priori assumed non-monotonic effect of task automation, but results from the interaction between new research opportunities and competition.

\begin{table}[htbp!]
    \centering
    \begin{tabular}{c l c}
        \toprule
        Parameter & Description & Baseline Value \\
        \midrule
        $\alpha$ & Share of AI-automated tasks & (0,1)\\
        $\eta$     & Elasticity of knowledge gain $G(d)$ & 0.6 \\
        $\theta$ & Proportional marginal increase in opportunities $\psi'(d)/\psi(d)$ & 0.5 \\
        $\delta$  & Congestion elasticity $N\xi'(N) / \xi(N)$ & 0.5\\
        $r$ & Discount rate & 0.05\\
        $a$ & Increase in recombinant capability thanks to AI & 2.0\\
        $c$ & AI cost relative to labor & 0.2\\
        $\beta$ & Baseline difficulty in recombination due to distance & 1.0\\
        $\epsilon$ & Final good elasticity with respect to intermediate good & 0.93\\
        $\bar{L}$ & Total labor supply & 1.0\\ 
        $p_0$ & Baseline discovery arrival at $d=0$ & 0.1\\
        $\lambda_0$ & Recombination capability without AI & 1.0\\
        \bottomrule
    \end{tabular}
    \caption{Parameters used in the numerical simulation.}
    \label{tab:simulation_parameters}
\end{table}

\begin{figure}
    \centering
    \includegraphics[width=\linewidth]{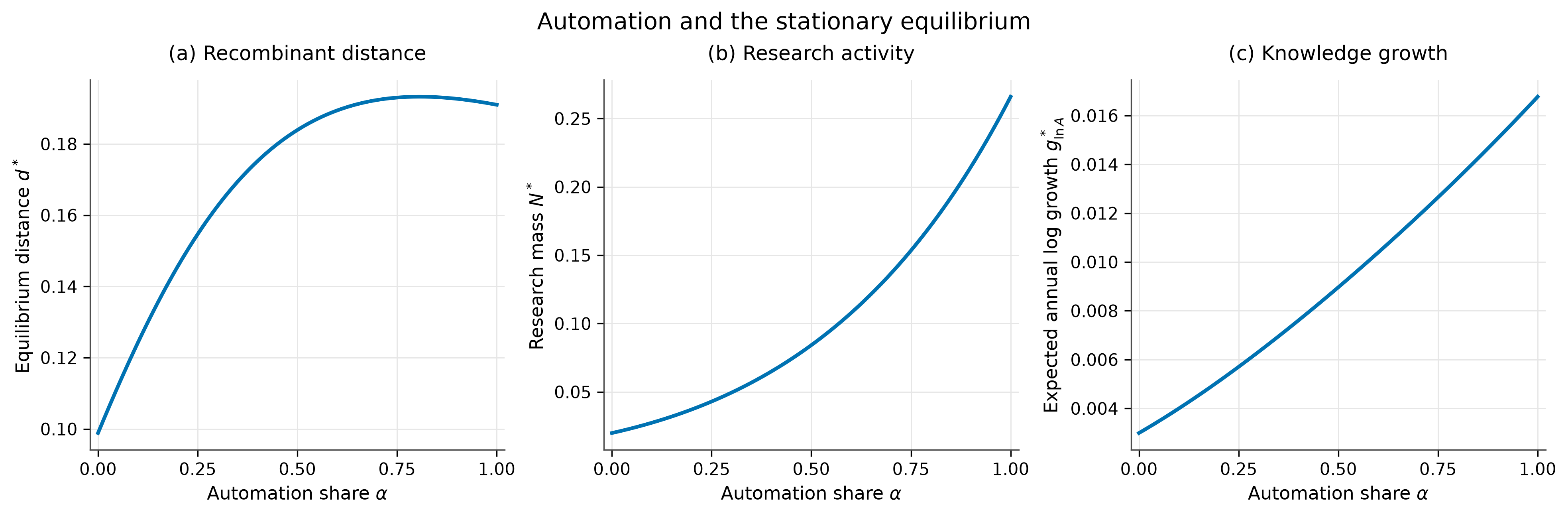}
    \caption{Baseline numerical simulation of the model with non-monotone behavior of equilibrium distance.}
    \label{fig:baseline_sim}
\end{figure}

Fig. \ref{fig:baseline_sim} shows stationary equilibrium distance and research mass, as well as the expected annual growth rate as the share of tasks performed by AI increases. We see that distance peaks around $\alpha\approx 0.80$, then starts declining: after that value, researchers focus more on exploitation of ideas in the knowledge space, rather than exploring more distant combinations. By converse, knowledge continues to grow as result of recombination facilitation and higher innovation arrival due to a ever-increasing mass of research units. Therefore, even if an increasing share of AI-automated tasks favors explorative combinations which are, conditional on success, more profitable, the joint effect of new opened up opportunities for subsequent researchers and competition, may revert research on more exploitative combinations. 

\begin{figure}
    \centering
    \includegraphics[width=\linewidth]{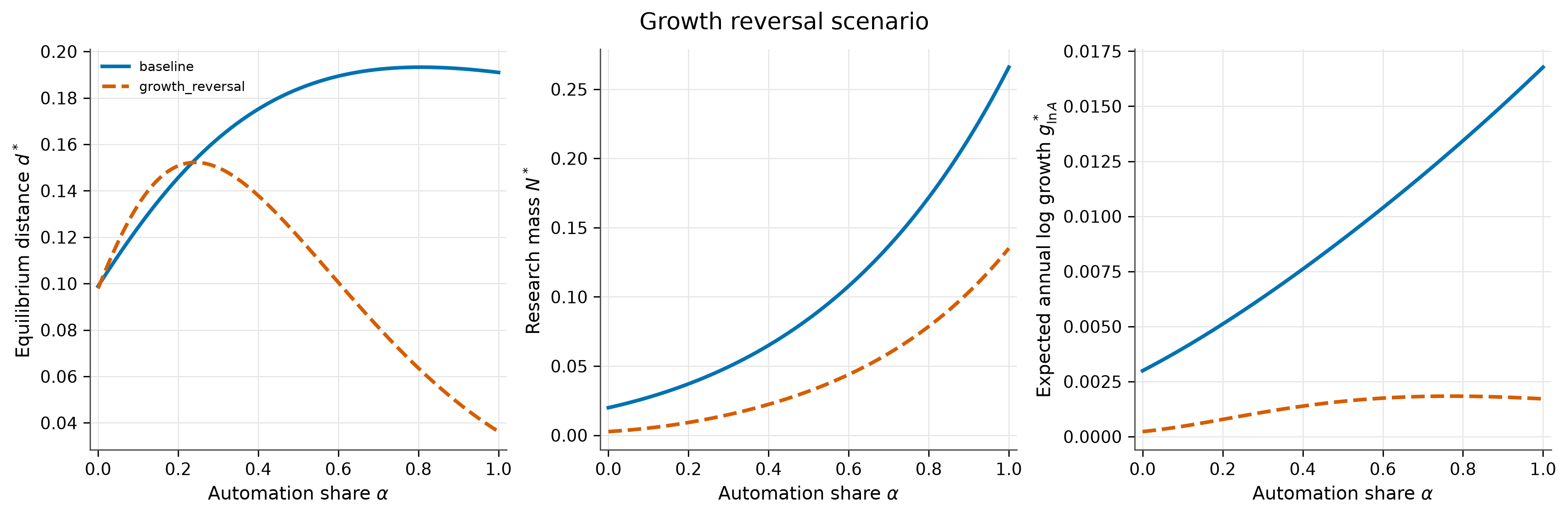}
    \caption{Numerical simulation of the model with slowing growth.}
    \label{fig:growth_reversal}
\end{figure}

Fig. \ref{fig:growth_reversal} shows an alternative parameter configuration with $(\eta,\theta,\delta)=(0.8,1,0.3)$. With this parameter configuration distant innovations generate larger knowledge gains but also open up more opportunities for subsequent research. Because congestion is relatively weak, new entry strongly increases the pace of discovery, making it easier for competitors to displace the incumbent. Equilibrium distance peaks at $\alpha \approx 0.25$, then as automation expands, firms respond to this threat by choosing shorter distances but the expected annual log-growth of knowledge continues to increase even with decreasing distance and peaks around $\alpha \approx 0.7$ because more frequent discoveries compensate for the smaller knowledge gain from each innovation. Indeed, we saw from eq. \eqref{cs_calendar_growth} that the peaks of the share of automated tasks for the equilibrium distance and knowledge growth need not to coincide. Beyond a certain level of automation, however, the combined contribution of research mass and the facilitation in combining ideas thanks to AI is not sufficient to offset the distance decrease and further AI-enabled research slows knowledge growth. 

The existence of a balanced growth path is supported by the assumption that AI price scales proportionally with the knowledge stock (Assumption 1). As already pointed out, a possible supply-side justification is that AI is produced competitively using labor with constant productivity $a_{AI}$. Assuming that labor employed in research and AI production is small relative to final good production allows us to neglect their feedback on the determination of equilibrium wage and therefore, we have $w = \iota A$. Competitive AI pricing gives $\mu(A) = w(A)/a_{AI}$.

In Appendix \ref{app:ai_pricing} we relax the assumption of constant productivity in AI production and numerically show optimal distance choice, research firm mass and expected knowledge log-growth along the the transition paths with varying relative AI price.

\section{Welfare Analysis}
We finally study whether along the balanced growth path, the privately chosen distance is optimal from a social point of view. We compare stationary allocations with a common initial knowledge stock $A_0$, holding automation and research mass fixed at $\alpha$ and $N^*$. In each candidate
allocation, the inherited distance equals the constant
research target $d$. This comparison does not evaluate
a transition from a common inherited state. Let $L_R^*=N^*l_R$ and $L_{AI}^*$ denote total labor
employed in research and AI production, respectively.

As described in the previous section, AI is produced competitively using labor with constant productivity, therefore the total quantity of AI supplied is:
\begin{equation}
    X_{AI}^S = a_{AI}L^*_{AI},
\end{equation}
and competitive pricing gives: $\mu=w/a_{AI}$. Total demand of AI services is given by:
\begin{equation}
    X_{AI}^D = \frac{\alpha A\Gamma N^*}{\mu}.
\end{equation}
AI-market clearing, $X_{AI}^S=X_{AI}^D$, together with
competitive pricing determines AI-production employment.
Cost minimization in research determines direct research
employment. These quantities are:
\begin{equation}
    L_{AI}^* = \frac{\alpha A\Gamma N^*}{w} \qquad L_R^* = \frac{(1-\alpha) A\Gamma N^*}{w}.
\end{equation}

Under the assumption that labor in research and AI production is small relative to total labor supply, we still retain $w = \iota A$ which implies:
\begin{equation}
    L_R^*+L_{AI}^* = \frac{\Gamma N^*}{\iota} \ll \bar{L}.
\end{equation}
Consequently from the final good production function \eqref{fin_good_prod}, within the maintained wage approximation, the consumption flow per unit of calendar time is:
\begin{equation}
    C(\nu) = A(\nu) \Omega^*, \qquad \Omega^* \equiv \left(\bar{L}- \frac{\Gamma N^*}{\iota}\right)^\epsilon.
\end{equation}

We evaluate stationary welfare using logarithmic utility,
assuming that the utility discount rate coincides with
the private valuation rate $r$:
\begin{equation}
    W(d,N^*) = \mathbb{E}\left[\int_0^\infty\exp(-r\nu) \ln C(\nu) d\nu \right] = \frac{\ln A_0+\ln\Omega^*}{r} + \frac{g(d, N^*)}{r^2},
    \label{welfare}
\end{equation}
where $g(d, N^*) = \mathcal{H}(d, N^*) \ln Q(d)$. We study welfare variation for local change in recombinant distance:
\begin{equation}
    \frac{\partial W(d, N^*)}{\partial d} = \frac{1}{r^2}\frac{\partial g(d, N^*)}{\partial d}.
    \label{der_welfare}
\end{equation}

Evaluating it at the decentralized stationary equilibrium pair gives:
\begin{equation}
    \left. \frac{\partial W}{\partial d} \right|_{(d^*,N^*)} = \frac{g(d^*,N^*)}{r^2} \left[ u(d^*) + \tau^*\chi^* + h(d^*) \left( \frac{1}{\ln Q(d^*)} -1 \right) \right].
\end{equation}
Therefore, we have:
\begin{equation}
    \ln Q(d^*) <1 \implies \left. \frac{\partial W}{\partial d} \right|_{(d^*,N^*)} >0.
    \label{sufficient_condition}
\end{equation}

This sufficient condition implies that, when the proportional knowledge gain per discovery satisfies $\gamma(d^*)<\exp(1)-1$, the decentralized stationary
pair ($d^*, N^*$) is not locally welfare-maximizing among the stationary allocations considered. A planner maintaining the same research mass could increase expected discounted log-consumption utility through a sufficiently small increase in the common stationary
distance. The same condition makes the distance coefficient in \eqref{cs_growth_distance_coefficient} positive. Consequently, stationary expected log-growth also increases when research mass and automation are held fixed. 

Although private research firms appropriate larger profit flows conditional on success when targeting longer distances, they also consider the shorter monopoly duration following from the greater subsequent research opportunities opened up by greater exploration of the knowledge space. However, they do not internalize the future knowledge and consumption gains from future discoveries and therefore, under the sufficient condition \eqref{sufficient_condition} research firms private choice yields a locally sub-optimal recombinant distance.

\subsection{By how much can the planner increase recombinant distance?}
Our sufficient condition \eqref{sufficient_condition} establishes that a sufficiently small increase above the decentralized equilibrium allows the planner to raise welfare, given a fixed level of automation and research mass. It remains to be studied by how much can the planner increase recombinant distance without decreasing social welfare. To answer this question, we exploit the functional forms (equation \eqref{functional_forms}) used in the simulation analysis.

From eq. \eqref{welfare}, we see that the consumption factor $\Omega^*$ is independent of distance and therefore, maximizing social welfare is equivalent to maximizing the expected log-growth $g(d, N^*)$. We define $q(\alpha):=\frac{\beta}{\lambda_0+\lambda\alpha}$ and rewrite eq. \eqref{der_welfare} as:

\begin{equation}
    \frac{\partial W(d,N^*)}{\partial d}
    =
    \frac{g(d,N^*)}{r^2}B(d;\alpha),
    \label{welfare_peak_derivative}
\end{equation}
where
\begin{equation}
    B(d;\alpha)
    :=
    \theta-q(\alpha)
    +
    \frac{\eta d^{\eta-1}}
    {(1+d^\eta)\ln(1+d^\eta)}.
    \label{welfare_peak_marginal}
\end{equation}

Thus, welfare increases with distance when $B(d;\alpha)>0$
and decreases when $B(d;\alpha)<0$. With $\eta \in (0,1)$, the last term of $B(d; \alpha)$, starts at infinite and strictly decreases to zero. Therefore, if $q(\alpha) > \theta$, there is a unique distance maximizing social welfare $d^P>0$, defined by:
\begin{equation}
    \frac{\eta(d^P)^{\eta-1}}
    {[1+(d^P)^\eta]\ln[1+(d^P)^\eta]}
    =
    q(\alpha)-\theta.
    \label{welfare_peak_foc}
\end{equation}

Under equation \eqref{sufficient_condition}, we therefore have that welfare increases for $d^*<d<d^P$ and decreases for $d>d^P>d^*$.

Finally, we measure the maximum increase in distance that can be chosen by the planner with respect to the decentralized optimal distance, defined as:
\begin{equation}
    \Delta_{\mathrm{PEAK}}
    :=
    d^P-d^*>0.
    \label{welfare_peak_distance_gap}
\end{equation}

From eq. \eqref{welfare_peak_foc}, we can find the following upper bound for the maximum increase in the planner's distance:
\begin{equation}
    0<\Delta_{\mathrm{PEAK}}
    <
    \frac{\eta}{q(\alpha)-\theta}-d^*.
    \label{welfare_peak_gap_bound}
\end{equation}

Notice that in order to have a finite social maximum, we need:
\begin{equation}
    q(\alpha)>\theta
    \quad\Longleftrightarrow\quad
    \alpha<
    \alpha_c
    :=
    \frac{\beta/\theta-\lambda_0}{\lambda},
    \label{welfare_peak_automation_threshold}
\end{equation}
for otherwise, the social welfare could be increased with no bounds. In words, we need that the discovery penalty associated with distance remains weaker than the opportunity spillover effects.

\section{Conclusions}
Knowledge advances through the recombination of previously existing ideas, in a process akin to the cross-pollination of seeds that produces new cultivated varieties \citep{weitzman1998}. The increasing integration of AI into R\&D activities provides researchers with powerful tools to navigate the broad space of possible idea combinations, identify promising candidates, generate hypotheses and accelerate experimentation. Yet, faster research does not necessarily mean more radical
research. The impact of AI on the nature of innovation, that is radical breakthroughs or incremental refinements, remains a central open question.
 
In this paper, we have addressed this question by developing a model of recombinant innovation embedded in a Schumpeterian quality-ladder framework. Ideas are located in a knowledge space where distances capture the cognitive dissimilarity between fields. R\&D firms choose how far to reach across this space, facing a fundamental trade-off: distant recombinations are riskier but lead to more radical and profitable innovations, while close recombinations are easier to develop but generate only incremental advances. Our distinctive contribution is to examine how competition changes this choice when AI becomes more widely adopted in the research process. AI makes distant recombinations relatively easier, but it also helps competitors innovate and can encourage additional research entry. As distant combinations open more opportunities for subsequent research potentially shortening monopolies, firms face the following dilemma: they can attempt more ambitious projects, but they may have less time to recover the rewards before being displaced. This connects the choice between exploration and exploitation to the competitive dynamics of creative destruction.

Our analysis points out three main results. First, an increasing fraction of research tasks performed by AI may have a non-monotonic effect on recombinant distance. Under the condition we have identified and as shown by the simulation, greater automation initially encourages exploration, but over-reliance on AI pushes firms back toward exploitative combinations. Such a reversal comes from competitive pressures due to faster replacement threat, rather than from an assumed decline in AI power, like the threat of novelty loss. Second, even with decreasing optimal distance along the balanced growth path, the overall expected log-growth of the knowledge stock may continue to increase thanks to a higher frequency of innovation and costs saving implied by AI. Our numerical simulation also shows that knowledge growth may slow down if distance falls fast enough. Third, under a sufficient condition on the knowledge gain per discovery, the decentralized distance is locally sub-optimal from a social point of view. Holding automation and research mass fixed, a planner could increase social welfare choosing a greater distance: under the private optimum, research is too incremental.

These findings contribute to the recent modeling literature on AI and economic growth and complements existing works on AI-assisted research and idea production. If on one hand, AI makes firms more productive, by lowering research costs and increasing their innovations arrival rate, on the other hand, it also changes the competitive environment by empowering all firms and accelerating the rate at which monopoly positions are displaced. This tension between individual benefits and aggregate displacement produces the usual exploration-exploitation trade-off.
 
Beyond the modeling contribution, our findings have broader implications for innovation policy and the governance of AI in research. The non-monotonic relationship between AI adoption and research novelty suggests that there may exist an optimal degree of AI integration in R\&D, which may exploit AI's capacity to bridge distant knowledge domains without sacrificing the creative diversity that human researchers provide. Policies that promote AI adoption in research should therefore carefully monitor and assess the risks of over-automation which, as described above, could also lead to a duplication of effort in scientific and R\&D research, and hence a slowdown in knowledge creation. This is in line with the recent empirical evidence showing that benefits of AI on scientific discovery critically depend on the judgment capabilities of human researchers as well as the risks of a reduction in the set of studied topics \citep{lou2021, tranchero2024a, hao2026artificial}. Such a reduction would be associated with a stronger exploitation of ideas in well-established domains for which we may already have high understanding and data abundance.

Furthermore, the role of monopoly duration in shaping firms' research direction has implications for intellectual property policy: our simulations suggest that regimes providing stronger and longer-lasting protection for innovators may encourage more radical recombinations, whereas environments with rapid displacement favor incrementalism. In an era where AI is accelerating the pace of innovation, the design of patent systems and the regulation of AI-assisted research necessitate careful reconsideration.
 
Our analysis has limitations that point to relevant directions for future research. First, the model treats AI as an exogenous input with a given price trajectory, without modeling the strategic behavior of AI providers. At present, the market for AI tools is quite concentrated, and the pricing, quality and accessibility of AI depend on the decisions of a small number of firms. Endogenizing AI provision, for instance, by introducing a monopolist or oligopolistic AI sector that sets prices and invests in capability improvements, would allow for a richer analysis of how the structure of the AI market affects the direction of innovation. \citet{gans2025growth} shows that monopolistic AI provision can have unexpected welfare benefits by restricting adoption; our framework could be extended to explore similar questions in the context of recombinant innovation and creative destruction. 

Second, our model introduces a graph-based representation of the knowledge space that we only use to define the notion of distance between pairs of ideas and build a simple analogy for the opportunity spillovers. However, our analysis does not exploit the network topology.
Future works could study how the structure of the knowledge graph influences the type of innovation that is developed: for instance, dense regions of the network where scientific understanding is already well-established may benefit most from long-distance recombinations that open entirely new research trajectories (knowledge expansion), while sparse, newly emerging areas may benefit more from incremental, short-distance combinations that consolidate and deepen existing knowledge (knowledge deepening). Furthermore, the graph representation opens promising avenues for empirical operationalization. For instance, knowledge distances can be measured using patent citation networks \citep{chen2017}, semantic similarity of scientific publications \citep{gentzkow2019}, or co-occurrence patterns in research output. Dynamic network analysis could track how successful recombinations create new connections and change subsequent opportunities, allowing the relationship between AI use, network structure, and research direction to be tested.

\clearpage

\bibliographystyle{apalike}
\bibliography{references.bib}

\section*{Funding Declaration}
The authors did not receive any financial support for this research.

\section*{Author contributions statement}

All authors contributed equally on idea conception, analysis and writing and revision of the paper.

\section*{Competing Interests}

The authors declare no competing interests.

\section*{Declaration of generative AI and AI-assisted technologies in the manuscript preparation process}

During the preparation of this work, the authors used
ChatGPT to assist with writing and debugging numerical
code, improving grammatical correctness and clarity,
and developing and checking mathematical derivations
and proofs. The authors reviewed, verified, and edited
the resulting material and take full responsibility
for the content of the published article. 

\clearpage

\appendix
\section{Appendix - Proofs}
\subsection{Replacement proportional response}
\label{subsec:replacement_change}
Consider one equilibrium with common research target $d^*$
and fixed parameters $\alpha,m$. Let $z$ denote a generic
inherited state. From eq. \eqref{continuation}, we have:
\begin{equation}
    \xi(N(z))=\psi(z)/k.
\end{equation}

Differentiating with respect to $z$, dividing by $\xi(N(z))$, and substituting \eqref{u} and \eqref{e} yields:
yields:
\begin{equation}
    \frac{N'(z)}{N(z)}
    =
    \frac{u(z)}{e(N(z))}.
    \label{eq:appendix_entry_response}
\end{equation}

Differentiating the aggregate Poisson arrival rate (eq. \eqref{continuation_intensity}) and dividing by $H(d^*\mid z)$  gives:
\begin{equation}
    \frac{1}{H(d^*\mid z)}
    \frac{\mathrm d}{\mathrm d z}H(d^*\mid z)
    =
    \frac{N'(z)}{N(z)}
    =
    \frac{u(z)}{e(N(z))}.
    \label{eq:inherited_state_response}
\end{equation}

\subsection{Proof of Proposition 1}
\label{proof:private}
\begin{proof}
Fix the exogenous parameters, the finite constant
$k>0$, and the finite subsequent target $\bar d>0$. Research firms maximize the logarithm of gross return \eqref{private_log_return} which does not depend on the current state $(A, z)$. Differentiating and substituting equations \eqref{h}, \eqref{s} and \eqref{eq:continuation_response} gives:
\begin{equation}
    L'(x)
    =
    h(x)-s(x)
    -
    \frac{u(x)}{e(N(x))}
    \frac{H(\bar d\mid x)}{r+H(\bar d\mid x)}.
    \label{proof_private_first}
\end{equation}
We now study the behavior of this equation at boundaries, as well as the sign of its derivative. The assumptions $\gamma(0)=0$ and
$\lim_{d\to 0^+}\gamma'(d)=+\infty$ imply
\begin{equation}
    h(x)
    =
    \frac{\gamma'(x)}{1+\gamma(x)}
    \longrightarrow+\infty
    \qquad\text{as }x\rightarrow0^+.
    \label{proof_private_zero_quality}
\end{equation}
The specified research-mass function satisfies
\begin{equation}
    \lim_{x\rightarrow0^+}N(x)
    =
    \xi^{-1}\left(\frac{\psi(x)}{k}\right)
    \in(0,\infty),
    \label{proof_private_zero_mass}
\end{equation}
because $\psi(0^+)$ and $k$ are finite and strictly
positive, and $\xi$ is continuous, strictly
increasing, and ranges from zero to infinity. Thus $e(N(x))$ converges to a strictly positive
value. Since $u(x)$ and $s(x)$ have finite
limits at zero and
\begin{equation}
    0<
    \frac{H(\bar d\mid x)}{r+H(\bar d\mid x)}
    <1,
    \label{proof_private_share_bound}
\end{equation}
the discovery and replacement penalties in
\eqref{proof_private_first} remain bounded near
zero. Therefore,
\begin{equation}
    L'(x)\longrightarrow+\infty
    \qquad\text{as }x\rightarrow0^+.
    \label{proof_private_zero_slope}
\end{equation}
For any $x>0$, concavity of $\gamma$ implies:
\begin{equation}
    \gamma(0)
    \leq
    \gamma(x)+\gamma'(x)(0-x).
    \label{proof_private_tangent_bound}
\end{equation}
Using $\gamma(0)=0$ gives
$x\gamma'(x)\leq\gamma(x)$. Hence:
\begin{equation}
    0<h(x)
    \leq
    \frac{1}{x}
    \frac{\gamma(x)}{1+\gamma(x)}
    \leq\frac{1}{x}
    \longrightarrow0
    \qquad\text{as }x\to\infty.
    \label{proof_private_large_quality}
\end{equation}

By assumption, $s(x)\geq\underline s>0$
at sufficiently large distances. Since the
replacement penalty in
\eqref{proof_private_first} is nonnegative,
\begin{equation}
    L'(x)
    \leq h(x)-s(x)
    \leq\frac{1}{x}-\underline s<0
    \qquad\text{for sufficiently large }x.
    \label{proof_private_large_slope}
\end{equation}

The derivative $L'$ is continuous and strictly
decreasing. It is positive near zero and negative
at sufficiently large distances. It remains to study the concavity of the objective function. Since $\gamma'>0$ and $\gamma''<0$ and $Q(x)=1+\gamma(x)$, we have:
\begin{equation}
    h'(x)
    =
    \frac{\gamma''(x)}{Q(x)}
    -
    \left[\frac{\gamma'(x)}{Q(x)}\right]^2
    =
    \frac{\gamma''(x)}{Q(x)}-h(x)^2
    \leq-h(x)^2<0.
    \label{proof_private_hprime}
\end{equation}
Differentiating the proportional replacement \eqref{eq:continuation_response} gives:
\begin{equation}
    \frac{\mathrm d}{\mathrm dx}
    \left[\frac{u(x)}{e(N(x))}\right]
    =
    \frac{u'(x)}{e(N(x))}
    -
    \frac{u(x)e'(N(x))N'(x)}
         {e(N(x))^2}
    \geq0,
    \label{proof_private_response_derivative}
\end{equation}
since $u'\geq0$, $u\geq0$, $e>0$, and
$e'\leq0$. Moreover, by $H'(x)=H(x) u(x)/e(N(x))$, we get:
\begin{equation}
    \frac{\mathrm d}{\mathrm dx}
    \left[
        \frac{H(\bar d\mid x)}{r+H(\bar d\mid x)}
    \right]
    =
    \frac{r}{[r+H(\bar d\mid x)]^2}
    \frac{\mathrm d}{\mathrm dx}H(\bar d\mid x)
    =
    \frac{u(x)}{e(N(x))}
    \frac{rH(\bar d\mid x)}
         {[r+H(\bar d\mid x)]^2}>0.
    \label{proof_private_share_derivative}
\end{equation}
Hence, differentiating equation \eqref{proof_private_first} gives:
\begin{equation}
\begin{aligned}
    L''(x)
    ={}&
    h'(x)-s'(x)\\
    &-
    \left[
        \frac{u'(x)}{e(N(x))}
        -
        \frac{u(x)e'(N(x))N'(x)}
             {e(N(x))^2}
    \right]
    \frac{H(\bar d\mid x)}{r+H(\bar d\mid x)}\\
    &-
    \left[\frac{u(x)}{e(N(x))}\right]^2
    \frac{rH(\bar d\mid x)}
         {[r+H(\bar d\mid x)]^2}
    <0.
\end{aligned}
    \label{proof_private_second}
\end{equation}
Consequently, $L$ is strictly concave on
$(0,\infty)$ and if a critical point exists, it is a maximum.
Together, we the study of the sign of $L'$ at boundaries, we conclude that there exists exactly one finite $x^*>0$ for
which $L'(x^*)=0$. Strict concavity ensures that $x^*$ is the unique interior global maximizer of $L$ and of $J(x\mid A,z)$.
\end{proof}

\subsection{Proof of Proposition 2}
\label{proof:existence}
Let $d>0$ and $N>0$ denote candidate values of stationary
distance and active research mass and let us define:
\begin{equation}
\begin{aligned}
    \rho(d,N)&:=\frac{\psi(d)p(d)}{\xi(N)}, \qquad
    \mathcal H(d,N):= N\rho(d,N),\qquad
    S(d):=\frac{\Pi Q(d)}{\Gamma},\\
    \tau(d,N)&:= \frac{u(d)}{e(N)}, \qquad
    \chi(d,N):=
    \frac{\mathcal H(d,N)}{r+\mathcal H(d,N)},
\end{aligned}
    \label{p2_definitions}
\end{equation}
where $\mathcal H(d,N)$ is the aggregate
arrival rate at a candidate stationary
allocation: the inherited distance and research
target both equal $d$, while $N$ remains an
unknown research mass. At the equilibrium,
$\mathcal H(d^*,N^*)=H(d^*\mid d^*)=H^*$.

We can rewrite the distance first-order condition \eqref{equilibrium_distance} and free-entry \eqref{stationary_free_entry} as:
\begin{align}
    F(d,N)
    &:=
    h(d)-s(d)-\tau(d,N)\chi(d,N)=0,
    \label{p2_F}\\
    E(d,N)
    &:=
    \rho(d,N)\bigl[S(d)-N\bigr]-r=0,
    \label{p2_E}
\end{align}
respectively.

\paragraph{Existence.}
\begin{proof}
First, we show that each proposed stationary distance $d>0$ determines a unique research mass satisfying free entry. Then, we show that at least one such pair also satisfies the distance first-order condition.

For fixed $d>0$, partial differentiation of the free-entry equation with respect to $N$, gives:
\begin{equation}
    E_N(d,N)
    =
    -\rho(d,N)
    -
    \frac{e(N)\rho(d,N)}{N}
    \bigl[S(d)-N\bigr]
    <0
    \qquad\text{for }0<N<S(d),
    \label{p2_EN_general}
\end{equation}
since $\rho_N(d,N)=-\frac{e(N)\rho(d,N)}{N}$. Thus $E(d,N)$ is strictly decreasing in active research mass on this interval. As $N\to0^+$, the numerator $\psi(d)p(d)[S(d)-N]$ approaches the strictly positive value $\psi(d)p(d)S(d)$, while
$\xi(N)\to0$. Therefore,
\begin{equation}
    \lim_{N\to0^+}E(d,N)=+\infty,
    \qquad
    E(d,S(d))=-r<0.
    \label{p2_entry_boundaries}
\end{equation}
Continuity and strict monotonicity of $E(d,N)$
in $N$ on $(0,S(d))$ imply that there is
exactly one research mass
\begin{equation}
    n_{\mathrm{FE}}(d)\in(0,S(d))
    \quad\text{such that}\quad
    E(d,n_{\mathrm{FE}}(d))=0.
    \label{p2_entry_curve}
\end{equation}

The function $n_{\mathrm{FE}}(d)$ gives the
research mass satisfying free entry at each
proposed stationary distance. Its continuity
follows from joint continuity of $E$ and its
strict decrease in $N$. 

Substituting the free-entry research mass into
the distance condition defines the continuous
function
\begin{equation}
    \Phi(d):=F(d,n_{\mathrm{FE}}(d)).
    \label{p2_phi}
\end{equation}

An equilibrium requires $\Phi(d)=0$.
We establish such a solution by showing that
$\Phi$ is positive near zero and negative at
sufficiently large distances. To determine the sign of $\Phi(d)$ near zero, we must first establish the boundary behavior of the
free-entry research mass $n_{\mathrm{FE}}(d)$.

For a fixed research mass $N\in(0,S_0)$, as $d\to0^+$ the free-entry residual gives:
\begin{equation}
    E(d,N)
    =
    \frac{\psi(d)p(d)}{\xi(N)}[S(d)-N]-r
    \longrightarrow f_0(N)-r,
    \label{p2_entry_residual_limit}
\end{equation}
where
\begin{equation}
    f_0(N):=\frac{\psi_0p_0}{\xi(N)}(S_0-N).
    \label{p2_f0}
\end{equation}

Since $f_0$ is continuous, $\lim_{N\to0^+}f_0(N)=+\infty$ , $\lim_{N\to S_0^-}f_0(N)=0$ and 
\begin{equation}
    f_0'(N)
    =
    -\frac{\psi_0p_0}{\xi(N)}
    -
    \frac{\psi_0p_0(S_0-N)\xi'(N)}
         {\xi(N)^2}
    <0,
    \label{p2_f0_derivative}
\end{equation}

then there is a unique value $N_0 \in (0, S_0)$ satisfying:
\begin{equation}
    \frac{\psi_0p_0}{\xi(N_0)}(S_0-N_0)=r.
    \label{p2_limiting_entry}
\end{equation}
It remains the check that $n_{\mathrm{FE}}(d)\to N_0$. Choose any two fixed masses $N\_$ and $N^+$ such that $0<N_-<N_0<N_+<S_0$. Since $f_0$ is strictly decreasing, we have $f_0(N_-)-r>0$ and $f_0(N_+)-r<0$. From eq. \eqref{p2_entry_residual_limit}, we have $E(d,N_-)>0$ and $E(d,N_+)<0$. Moreover, $N_+<S(d)$ for sufficiently small
$d$, since $S(d)\to S_0>N_+$. Thus, the unique research mass satisfying free entry must be such that $N_-<n_{\mathrm{FE}}(d)<N_+$ and as these bounds can be chosen arbitrarily close
to $N_0$:
\begin{equation}
    n_{\mathrm{FE}}(d)\longrightarrow N_0
    \qquad\text{as }d\to0^+.
    \label{p2_entry_zero_limit}
\end{equation}

Along the free-entry curve, we have:
\begin{equation}
    \Phi(d)
    =
    h(d)-s(d)
    -
    \frac{u(d)}{e(n_{\mathrm{FE}}(d))}
    \chi(d,n_{\mathrm{FE}}(d)),
    \label{p2_phi_expanded}
\end{equation}
where $e(n_{\mathrm{FE}}(d))\to e(N_0)>0$, thus
the denominator of $\tau(d,N)$ is bounded away from zero
for sufficiently small positive $d$. Together with the finite limit of $u(d)$ and $0<\chi<1$, this ensures that the
replacement penalty remains bounded. The quality-return term $h(d)$ diverges to $+\infty$ as $d\to0^+$, while both the discovery penalty $s(d)$ and the replacement penalty
remain bounded. Consequently,
\begin{equation}
    \Phi(d)\longrightarrow+\infty
    \qquad\text{as }d\to0^+.
    \label{p2_phi_zero}
\end{equation}
At large distances, the sign is reversed:
the concavity of $\gamma$ and $\gamma(0)=0$
imply $h(d)\leq1/d$, as shown in the proof
of Proposition 1. Since
$s(d)\geq\underline s>0$ at sufficiently
large distances and the replacement penalty
is nonnegative,
\begin{equation}
    \Phi(d)
    \leq h(d)-s(d)
    \leq\frac{1}{d}-\underline s<0
    \qquad\text{for sufficiently large }d.
    \label{p2_phi_large}
\end{equation}
There are two finite positive distances at which the continuous function $\Phi$ has opposite signs. By the intermediate value theorem, there exists a finite $d^*>0$ satisfying $\Phi(d^*)=0$. With $N^*=n_{\mathrm{FE}}(d^*)$, free entry holds by construction and the distance condition also holds establishing the existence of a positive solution to the stationary equilibrium system.
\end{proof}

\paragraph{Uniqueness.}
\begin{proof}
We now show that the positive stationary solution
is unique by showing that $\Phi'(d)<0$ at every zero
of $\Phi$. In the following, subscripts $d$ and $N$ denote partial derivatives and functions without displayed arguments are evaluated at $(d,N)$:
\begin{equation}
\begin{aligned}
    \rho_d
    &=\rho(u-s), \qquad
    \mathcal H_d=\mathcal H(u-s), \qquad
    \mathcal H_N=\frac{\mathcal H(1-e)}{N}, \qquad
    \tau_d=\frac{u'(d)}{e(N)},\\
    \tau_N&=-\frac{\tau e'(N)}{e(N)}, \qquad
    \chi_d=\chi(1-\chi)(u-s),\qquad
    \chi_N=\frac{\chi(1-\chi)(1-e)}{N}.
    \label{p2_chi_N}
\end{aligned}
\end{equation}
Partial differentiation of the distance condition \eqref{p2_F} yields:
\begin{align}
    F_d
    &=
    h'(d)-s'(d)
    -\frac{u'(d)}{e(N)}\chi
    -\tau\chi(1-\chi)(u-s),
    \label{p2_Fd}\\
    F_N
    &=
    \tau\chi
    \left[
        \frac{e'(N)}{e(N)}
        -
        \frac{(1-\chi)(1-e(N))}{N}
    \right]
    \leq0.
    \label{p2_FN}
\end{align}

The sign of $F_N$ follows from
$\tau\geq0$, $0<\chi<1$, $e'(N)\leq0$,
and $0<e(N)<1$. To determine the sign of $F_d$, consider any
joint solution of $F=0$ and $E=0$. The distance condition implies that $h=s+\tau\chi$. Also, concavity of $\gamma$ gives
$-h'\geq h^2$, as established in Proposition 1 and together with our assumptions $s'\geq0$, $u'\geq0$, and $u\geq0$, we get:
\begin{equation}
\begin{aligned}
    -F_d
    =
    -h'+s'
    +\frac{u'}{e}\chi
    +\tau\chi(1-\chi)(u-s)\geq
    h^2-\tau\chi(1-\chi)s
    &=
    (s+\tau\chi)^2-\tau\chi(1-\chi)s\\
    &=
    s^2+s\tau\chi(1+\chi)+(\tau\chi)^2
    >0.
\end{aligned}
    \label{p2_Fd_negative}
\end{equation}
Thus $F_d<0$ at every joint solution.

Partial differentiation of the free-entry condition \eqref{p2_E} yields:
\begin{equation}
    E_d
    =
    \rho(u-s)(S-N)+\rho Sh,
    \label{p2_Ed_general}
\end{equation}
since $S'(d)=S(d)h(d)$. At a joint solution, free entry implies
$\rho(S-N)=r$ and $\rho S=r+\mathcal H$, while distance condition implies $h=s+\tau\chi$. Combining them gives:
\begin{equation}
\begin{aligned}
    E_d
    =
    r(u-s)+(r+\mathcal H)h
    =
    r(u-s)+(r+\mathcal H)(s+\tau\chi)
    =
    ru+\mathcal H(s+\tau)>0.
\end{aligned}
    \label{p2_Ed_positive}
\end{equation}
The last equality uses
$(r+\mathcal H)\chi=\mathcal H$,
and the strict inequality follows from
$\mathcal H>0$ and $s>0$. Furthermore, substituting free entry into
\eqref{p2_EN_general} gives: 
\begin{equation}
    E_N
    =
    -\rho-\frac{er}{N}
    =
    -\frac{\mathcal H+er}{N}<0.
    \label{p2_EN_negative}
\end{equation}
Since $E$ is continuously differentiable and
$E_N<0$ at its free-entry roots, the implicit
function theorem gives
\begin{equation}
    n_{\mathrm{FE}}'(d)
    =
    -
    \frac{E_d(d,n_{\mathrm{FE}}(d))}
         {E_N(d,n_{\mathrm{FE}}(d))}.
    \label{p2_entry_curve_derivative}
\end{equation}
At every zero of $\Phi$, both equilibrium
conditions hold. Hence
\eqref{p2_Ed_positive} and
\eqref{p2_EN_negative} imply
$n_{\mathrm{FE}}'(d)>0$ at such a point.

The total derivative of the distance residual
along the free-entry curve is therefore
\begin{equation}
\begin{aligned}
    \Phi'(d)
    &=
    F_d(d,n_{\mathrm{FE}}(d))
    +
    F_N(d,n_{\mathrm{FE}}(d))
    n_{\mathrm{FE}}'(d)\\
    &<0
    \qquad\text{at every zero of }\Phi,
\end{aligned}
    \label{p2_phi_derivative}
\end{equation}
since $F_d<0$, $F_N\leq0$, and
$n_{\mathrm{FE}}'(d)>0$.

Thus every zero of $\Phi$ is a strict downward crossing: $\Phi$ is positive to its left and negative to its right. Two such zeros cannot coexist, since returning from negative to positive values between them
would require another zero that is not a strict downward crossing. This establishes exactly one zero $d^*>0$. Since each proposed distance determines exactly one free-entry research mass, the stationary system has a unique positive solution $(d^*,N^*)$, where $N^*=n_{\mathrm{FE}}(d^*)$. By Proposition 1, the private first-order condition is sufficient for unique global private optimality under the associated continuation activity. Hence this pair is the unique positive stationary equilibrium.
\end{proof}

\subsection{Proof of distance and research firm mass response}
\label{proof:comparative_statics}
\begin{proof}
Since $\psi$, $\xi$, and $Q$ do not directly
depend on $\alpha$, the definitions in
\eqref{cs_direct_effects} imply
\begin{equation}
    \rho_\alpha=\rho\ell_\alpha,
    \qquad
    \mathcal H_\alpha=\mathcal H\ell_\alpha,
    \qquad
    S_\alpha=-Sc_\alpha,
    \qquad
    \tau_\alpha=0,
    \label{cs_proof_direct}
\end{equation}
and
\begin{equation}
    \chi_\alpha
    =
    \frac{r\mathcal H_\alpha}
         {(r+\mathcal H)^2}
    =
    \chi(1-\chi)\ell_\alpha.
    \label{cs_proof_chi}
\end{equation}
Partial differentiation of equations \eqref{cs_F} and \eqref{cs_E} gives:
\begin{align}
    F_\alpha
    &=
    -s_\alpha-\tau\chi(1-\chi)\ell_\alpha.\\
    E_\alpha
    &=
    \rho\ell_\alpha(S-N)-\rho Sc_\alpha.
\end{align}

At equilibrium, $\rho(S-N)=r$ and
$\rho S=r+H^*$, thus:
\begin{equation}
    E_\alpha
    =
    r\ell_\alpha-(r+H^*)c_\alpha.
    \label{cs_proof_Ealpha}
\end{equation}
The Jacobian with respect to $(d,N)$ is
\begin{equation}
    \begin{pmatrix}
        F_d & F_N\\
        E_d & E_N
    \end{pmatrix}
    =
    \begin{pmatrix}
        -\mathcal A & -\mathcal B\\
        \mathcal C & -\mathcal D
    \end{pmatrix},
    \label{cs_proof_jacobian}
\end{equation}
whose determinant is
$\Delta=\mathcal A\mathcal D+
\mathcal B\mathcal C>0$. The implicit function theorem therefore
establishes local continuous differentiability of the equilibrium pair in $\alpha$.

Taking total derivatives of $F(d^*(\alpha),N^*(\alpha);\alpha)=0$
and $E(d^*(\alpha),N^*(\alpha);\alpha)=0$ gives:
\begin{align}
    \mathcal A\frac{\mathrm d d^*}{\mathrm d\alpha}
    +
    \mathcal B\frac{\mathrm d N^*}{\mathrm d\alpha}
    &=F_\alpha,
    \label{cs_proof_linear_F}\\
    -\mathcal C\frac{\mathrm d d^*}{\mathrm d\alpha}
    +
    \mathcal D\frac{\mathrm d N^*}{\mathrm d\alpha}
    &=E_\alpha.
    \label{cs_proof_linear_E}
\end{align}
Solving this linear system yields
\eqref{cs_distance_response} and
\eqref{cs_mass_response}.
\end{proof}

\section{Robustness checks for the numerical analysis}
\label{robust_app}
\begin{figure}[htbp!]
    \centering
    \includegraphics[width=\linewidth]{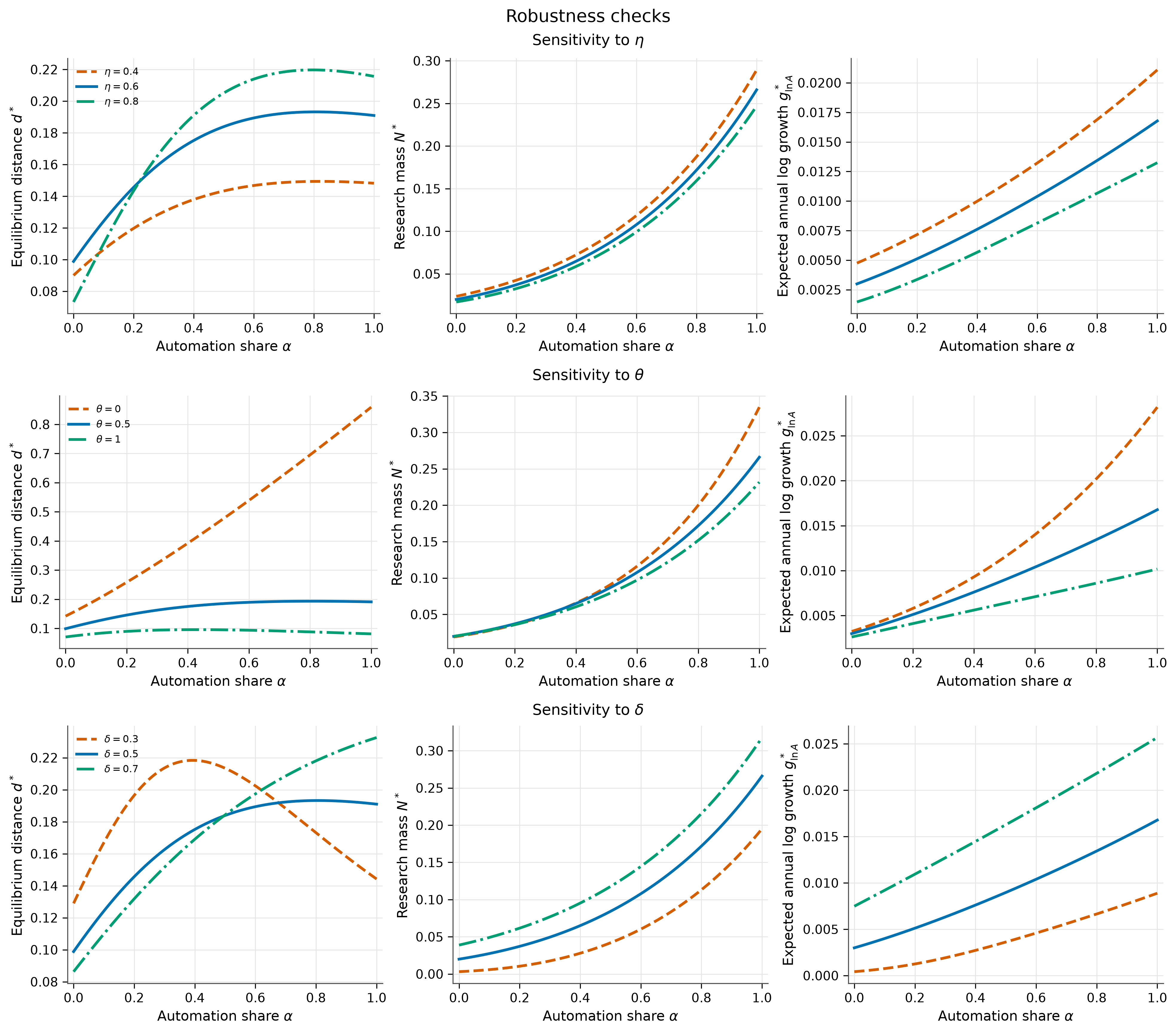}
    \caption{Robustness checks for the numerical simulation of the model with different values of $\eta\in\{0.4,0.6,0.8\}$, $\theta\in\{0,0.5,1\}$, and $\delta\in\{0.3,0.5,0.7\}$. These parameters are changed one at time, while all the others are kept fixed.}
    \label{fig:robustness}
\end{figure}

Fig. \ref{fig:robustness} shows equilibrium distance, research mass and expected annual log-growth of knowledge for different values of $\eta, \theta$ and $\delta$. We solve the model varying $\eta\in\{0.4,0.6,0.8\}$, $\theta\in\{0,0.5,1\}$, and $\delta\in\{0.3,0.5,0.7\}$ one at a time, while the other parameters are kept fixed. This exercise provides illustrative comparison rather than an empirical calibration of the model's parameters. 

Non-monotonicity of the equilibrium distance is preserved in almost all cases, except for two cases: (i) opportunities for subsequent research are distance-independent $\theta=0$, and (ii) high congestion elasticity $\delta$. In the first case, current combinations do not affect future innovation arrivals and therefore current researchers are not discouraged in targeting long distances as they do not face an increased future replacement rate ensuing their choice. In the second case, higher congestion decreases the overall innovation arrival rate and hence the incentive to target longer distances is not offset by a lower monopoly duration.

\clearpage
\section{Alternative AI-pricing rules}
\label{app:ai_pricing}

Let $A_0>0$ denote the initial level of knowledge and $\sigma=\ln(A/A_0)$ the logarithm of the ratio between current  and initial knowledge stocks.

We define the relative AI price:
\begin{equation}
    c(\sigma) := \frac{\mu(A)}{mw(A)} = c_\infty + (c_0-c_\infty)\exp(-j\sigma),
\end{equation}
with $0<c_0<1$ and $0<c_\infty \le 1$ being the relative AI price at the initial knowledge level and its limiting price as knowledge accumulates, respectively. The parameter $j>0$ determines how quickly the relative cost approach the limiting value and $c(y)<1$, in line with the technology adoption condition \eqref{tech_adop_cond}. 
If $c_0>c_\infty$, we have a declining relative AI price, whereas if $c_0<c_\infty$ the relative price is increasing with $A$.

We relax the constant productivity assumption in production of AI and numerically examines the transitions paths of optimal distance, research units mass and expected log-growth of knowledge with relative AI price either increasing or decreasing as knowledge accumulates. 

In this case AI relative price is given by $\mu(A)=m\iota Ac(\sigma)$ and research costs now depend on knowledge stock: $\Gamma(\sigma)=\iota (c(\sigma))^\alpha$. In the numerical solution, for each pricing scenario, we use as limiting boundary condition the stationary equilibrium associated with that scenario’s limiting relative AI cost $c_\infty$. Along the transition, however, research firms must account also for the subsequent AI price determined by the new level of knowledge stock in case their recombination is successful. The new price therefore affects future research costs and entry, as well as replacement intensity. 

\begin{figure}[htbp!]
    \centering
    \includegraphics[width=\linewidth]{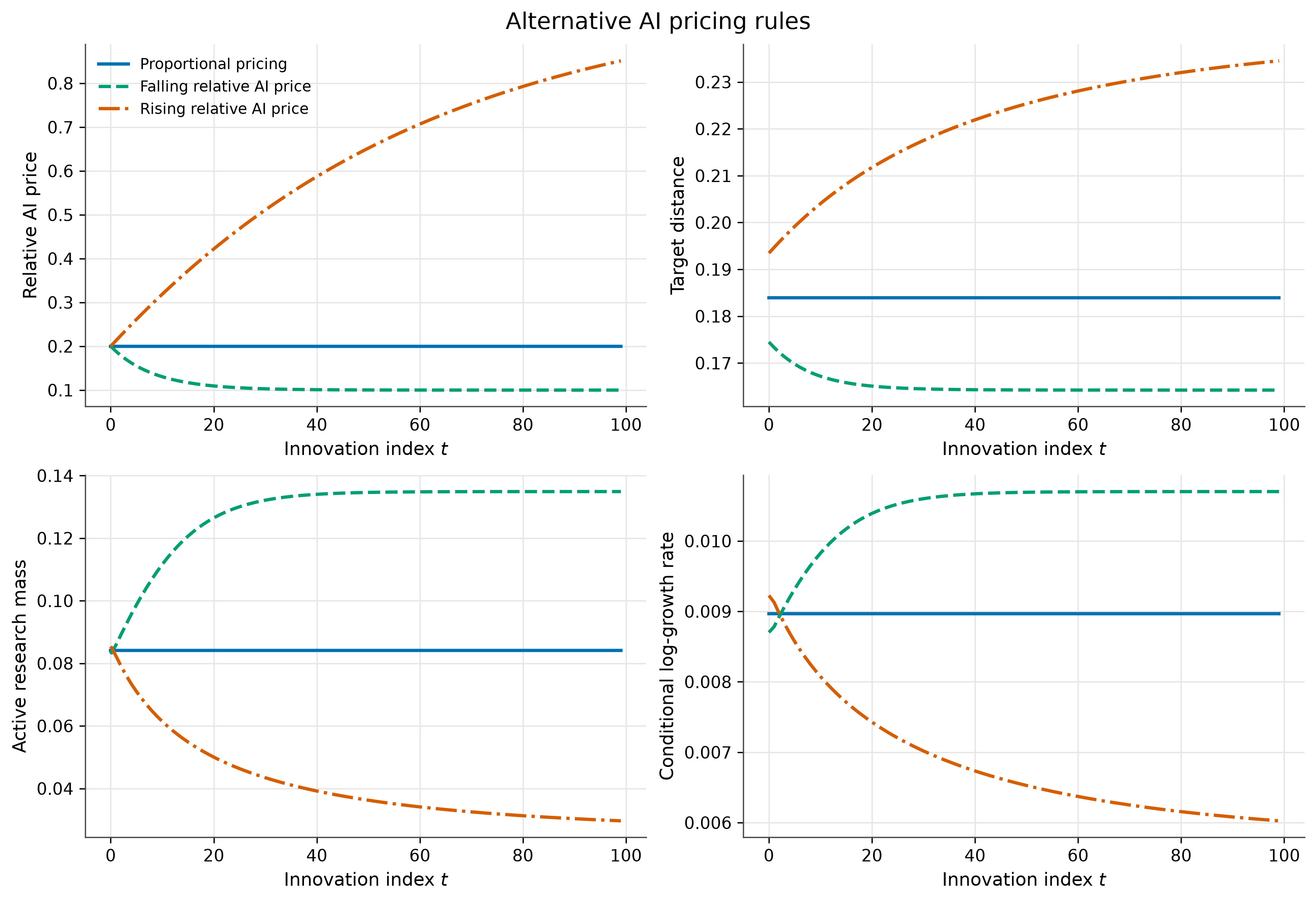}
    \caption{Alternative relative pricing rule for AI services.}
    \label{fig:alternative_ai_pricing}
\end{figure}

Figure~\ref{fig:alternative_ai_pricing} compares the
stationary proportional pricing benchmark with
transition paths under alternative AI-pricing rules. We retain the parameters' values adopted for the baseline model (see Table \ref{tab:simulation_parameters}) and we set $\alpha=0.5$. Further, $c_0=0.2$, $c_\infty=\{0.2,0.1, 1\}$ for the proportional, decreasing and increasing relative price scenario, respectively. 

Wee see that with a decreasing relative AI price, the optimal recombinant distance is lower than in the BGP case, but the reduction in research costs push research firms to enter the market. Relatively cheaper AI pushes pushes research toward more exploitative combinations, but results in a higher expected log-growth rate due to higher frequency of innovations. By converse, an increasing relative price of AI increases research costs and deter new firms to enter in the market. The ensuing lower replacement rate increases expected monopoly duration and favor riskier, more distant recombinations. However, the overall log-growth rate decrease over time due to the lower innovation frequency.

\end{document}